\documentclass[11pt]{article}
\begin{document}
\newcommand{\be}{\begin{equation}}
\newcommand{\ee}{\end{equation}}
\begin{center}
{\bf A Gas-Kinetic Stability Analysis of Self-Gravitating and
Collisional Particulate Disks with Application to Saturn's Rings} 
\end{center}

\begin{center}
{\bf Evgeny Griv$^{\ddag \S}$\footnote{Author to whom correspondence 
should be addressed at Beer-Sheva (email: griv@bgumail.bgu.ac.il).},
Michael Gedalin$^\ddag$, David Eichler$^\ddag$, and Chi Yuan$^\S$}\\
$^\ddag${\small Department of Physics, Ben-Gurion University 
of the Negev, P.O. Box 653, Beer-Sheva 84105, Israel}\\
$^\S${\small Academia Sinica Institute of Astronomy and
Astrophysics, P.O. Box 1-87, Taipei 11529, Taiwan}
\end{center}

   \begin{abstract}

Linear theory is used to determine the stability of the
self-gravitating, rapidly (and nonuniformly) rotating, two-dimensional,
and collisional particulate disk against small-amplitude gravity
perturbations.  A gas-kinetic theory approach is used by exploring the
combined system of the Boltzmann and the Poisson equations.
The effects of physical collisions between particles are taken into 
account by using in the Boltzmann kinetic equation a Krook model integral  
of collisions modified to allow collisions to be inelastic.  It is shown
that as a direct result of the classical Jeans instability and a
secular dissipative-type instability of 
small-amplitude gravity disturbances (e.g. 
those produced by a spontaneous perturbation and/or a companion system) 
the disk is subdivided into numerous irregular ringlets, with size and 
spacing of the order of $4 \pi \rho \approx 2 \pi h$, where $\rho \approx 
c_r / \kappa$ is the mean epicyclic radius, 
$c_r$ is the radial dispersion of random
velocities of particles, $\kappa$ is the local epicyclic frequency, and
$h \approx 2 \rho$ is the typical thickness of the system.  The present
research is aimed above all at explaining the origin of various structures 
in highly flattened, rapidly rotating systems of mutually gravitating
particles.  In particular, it is suggested that forthcoming Cassini
spacecraft high-resolution images may reveal this kind of hyperfine
$\sim 2 \pi h \stackrel{<}{\sim} 100$ m structure in the main rings 
A, B, and C of the Saturnian ring system.

   \end{abstract}


\section{Introduction}

Self-gravitating disk systems are of great interest in
astrophysics because of their widespread appearance, e.g. disks in
spiral galaxies, pancakes and accretion disks around massive objects,
low mass X-ray binaries, the protoplanetary clouds, and, finally, the 
main rings of Saturn.  Such systems are highly dynamic
and are subject to various instabilities of small-amplitude gravity 
perturbations.\footnote{The strongly flattened, disk-shaped form of all
these objects is due to their rapid rotation.  Thus, equilibrium is 
established in a simple manner in such disks, i.e. it is governed mainly
by the balance between the centrifugal and gravitational forces.} 
This is because the evolution of these disks is primarily driven by
angular momentum redistribution.  This might take place through global
mechanisms like (1) nonaxisymmetric instabilities caused by 
self-gravity or (2) instabilities caused by simultaneous 
action of self-gravity and effective viscous coupling between 
neighbouring disk annuli.

Because of the long-range nature of the gravitational forces
between particles, a self-gravitating system exhibits collective modes of
motions --- modes in which the particles in large regions move coherently
or in unison.  In turn, collective-type oscillations have been studied in
a system of electrically charged particles in a plasma (Alexandrov {\it et
al.}, 1984; Krall and Trivelpiece, 1986; Swanson, 1989).  
The similarity between the 
gravitational and Coulomb interactions is well known and already explored,
e.g. Bertin (1980), Fridman and Polyachenko (1984), and Binney and Tremaine
(1987).  Therefore, in this paper we employ certain developed mathematical
formalisms from plasma physics.

The dynamics of flattened gravitating systems has now been
studied quite thoroughly.  This research has aimed to explain the origin
of various observed structures: spiral and ring formations in flat galaxies,
the thin ringlets around Saturn, etc.  One of the main trends has therefore
been to analyze the perturbation dynamics in such systems, in both linear
and nonlinear regimes (Fridman and Polyachenko, 1984; Binney and
Tremaine, 1987).  

In the current research, the linear stability theory of small oscillations
(and their stability) of a disk of mutually gravitating particles is
reexamined by using the method of gas-kinetic theory which incorporates
interparticle collisions through a Krook (or the so-called
$\tau$-) approximation (Lifshitz and Pitaevskii, 1981; Krall and
Trivelpiece, 1986, p. 315).  This representation gives new insight
into the gravitating disk stability.  In particular, 
we will investigate in detail the important limit of strong 
and frequent interparticle collisions.  The results of the analysis are
applied to Saturn's rings, composed of rock and ice: 
we predict the small-scale, hyperfine structure 
of order $100$ m to be observed in the main rings A, B, and C.\footnote{
Saturn is, of course, most famous for its spectacular set of rings.
Attempts to find a plausible naturalistic explanation of the origin of
Saturn's rings began about 250 years ago but have not yet been quantitatively
successful, making this one of the oldest unsolved problems in modern science.
Many of famous mechanics and mathematics investigators, starting with Laplace
and Maxwell, studied their structure, composition, and stability.}  

Saturnian ring system populated primarily by centimeter- to a several 
meter-sized mutually gravitating and physically colliding particles
(Zebker {\it et al.}, 1985).  A particle size distribution function
exhibits approximately inverse-cubic power-law behavior.  In addition,
Saturn has extensive but much more tenuous rings containing mainly
micrometer-sized particles (Cuzzi {\it et al.}, 1984; Lissauer and Cuzzi,
1985).  Since collisions tend to dissipate the highly ordered motions involved 
in wave propagation, collisions will lead to wave dissipation.  Except for
rings of very small collision frequency, impacts are certain to dominate over
individual gravitational encounters.  Some of the properties of inelastic
physical collisions in Saturn's main rings are discussed in the article.

In fact, it should be made clear right from the start that the 
suggestion of hyperfine $\stackrel{<}{\sim} 100$ m
structure in Saturn's rings due to the combined effects of
gravity and interparticle collisions is not an entirely new idea. 
Salo (1992, 1995), Willerding (1992), Richardson (1994), Osterbart
and Willerding (1995), and Sterzik {\it et al.} (1995) have already
predicted such a structure in Saturn's rings by $N$-body computer
simulations.  Our significant contribution is just a kinetic theory 
derivation of results obtained before in simulations by Salo,
Osterbart, Willerding, Richardson, and Sterzik {\it et al.} (see also 
Griv (1998)) or through other analytic approximations (Fridman and 
Polyachenko, 1984, Vol. 2; Schmit and Tscharnuter, 1995, 1999).

In turn, the fairly recent 
Voyager missions have shown that Saturn's main rings (the
brightest A and B rings, and the semitransparent inner C ring) are
divided into a huge number of irregular concentric rings (Smith
{\it et al.}, 1982); see Cuzzi {\it et al.} (1984), Lissauer and Cuzzi
(1985), Esposito (1986, 1993), Cuzzi (1989),
and Nicholson and Dones (1991) as reviews.
The Voyager 2 spacecraft close-up view of Saturn's rings shows that
even the so-called gaps demonstrate a complicate structure ---
Cassini's division, for example, contains perhaps 100 ringlets (e.g.
Flynn and Cuzzi, 1989). 
It was found that the main rings exhibit large irregular
variations in optical depth that are not associated with any resonances
with known satellites.  Actually, a new class of objects in the solar
system was discovered.  Moreover, the planetary rings turned out to be 
a necessary element and consistent phenomenon in the satellite systems
of all giant planets.  
Because low-velocity collisions of ice particles will always involve
some dissipation of acoustic energy, some source of energy must be
available to keep them from total collapse to a featureless monolayer.

The best analysis of the Voyager 2  
photopolarimeter PPS data and imaging science data for structure was
done by Showalter and Nicholson (1990) and Horne and Cuzzi (1996),
respectively.  The Voyager 2 PPS experiment obtained the highest
resolution of any ring observation of Saturn, profiling the variation
of optical depth in radial steps of about 100 m.  However, below a few
kilometers scale, the PPS data is too noisy to extract information
about irregular structure: the finest ``structure" observed by PPS is
well fit by models of statistical noise combined with stochastic 
variations resulting from large particles or clumps of particles 
(Showalter and Nicholson, 1990).  So, such hyperfine $\sim 100$ m
structure might exist, but the Voyager 2 spacecraft could
not see it.  Imaging science data were sensitive only down
to $10-20$ km scales.  Some ring regions only exhibit well-defined
characteristics scales larger than that, while others do show power
on scales all the way down to the $10-20$ km lower limit (Horne and
Cuzzi, 1996).  Kilometer-scale and larger irregular structure is
primarily confined to Saturn's high optical depth B ring, where
particles collide frequently.  It is clear, however, that despite the
lack of evidence for the types of $\stackrel{<}{\sim} 100$ m ring structure
that we are trying to explain, neither is there evidence that such
structure does not exist; we simply do not have observations with
resolution comparable to the scale height of the main rings.

As was stated by Wisdom and Tremaine (1988), the presence of the irregular
structure in the Saturnian ring system is surprising because the timescale
on which such irregularities in the distribution of particles should be 
removed by viscous diffusion is much shorter than the age of the solar 
system.

In summary, the Voyager space missions have provided accurate data 
regarding the dynamics of planetary rings.
The Voyager 2 flyby of Saturn has revealed
that the Saturnian disk shows a complex irregular density structure
ranging from a few kilometers down to the several hundred meters'
resolution of the spacecraft's camera (Lane {\it et al.}, 1982;
Smith {\it et al.}, 1982; Cuzzi {\it et al.}, 1984; Esposito, 1986, 
1993; Sicardy and Brahic, 1990).
Most irregular microstructures are likely much
younger than the solar system and new rings are created by some unknown
mechanisms (Esposito, 1986, 1993).  For instance, Esposito (1986, 1993)
advocates the hypothesis of a larger role for catastrophic events in the
Saturnian rings: new rings are episodically created by the destruction 
of small moons near the planet.  Rings arise from singular events like
the destruction of a ringmoon or comet, and a ring's physical nature is 
the result of a competition between fragmentation and accretion in the
planet's Roche zone.  These processes often involve occasional major 
events. Ring history is disorderly.  Dones (1991)
even proposed a recent cometary origin for Saturn's rings.  Accordingly,
an origin for Saturn's rings in tidal disruption by a comet of the
scale of Chiron, which passed within Saturn's Roche radius, is explored.
Shan and Goertz (1991) suggested that the electromagnetically induced 
radial transport of angular momentum associated with radial transport of
charged submicron-size dust particles may explain the features in the
Saturnian B ring.  This mechanism induces an instability which produces, 
over geological times, significant radial structuring of the ring.
The radial scale of this irregular structure extends to very small 
sizes, down to the resolution limits.  As was stated by Esposito (1986):
``every time we improve our resolution we see more structue."

On a small scale the irregular rings have been observed to
be undergoing variation and oscillations with time and ring longitude
(Smith {\it et al.}, 1982).  Actually, it was found that the individual 
rings of Saturn are in various states of oscillations.
The latter indicates that probably such
features are wave phenomena, and different instabilities of gravity
perturbations may play important roles in ring's dynamics.  In this regard,
the wealth of ring data from the Voyager spacecraft already motivated 
studies of wave propagation in planetary rings (see Goldreich and
Tremaine (1982), Shu {\it et al.} (1983), Borderies (1989), Araki (1991),
and Nicholson and Dones (1991) as reviews of
the problem).\footnote{The rings of Saturn consist of thousands of smaller
ringlets.  In this paper, however, we do not consider few truly isolated
ringlets with adjacent empty gaps, located in the low-density C-ring and
the Cassini Division resembling those of Uranus (Tyler {\it et al.}, 1983; 
Porco and Nicholson, 1987; Porco, 1990).  Many of these narrow
ringlets (with typical widths of a few tens of kilometers) with 
extremely sharp edges are found in the isolated resonance locations of
different satellites, e.g. Prometheus $2:1$ inner Lindblad resonance.
An adequate theoretical explanation for these isolated narrow ringlets is
still missing (Hanninen and Salo, 1995; Goldreich {\it et al.}, 1995).} 
This includes externally driven satellite resonances which clear gaps
by perturbing ring particles at certain radial distances (Holberg {\it 
et al.}, 1982; Thiessenhusen {\it et al.}, 1995; Horn {\it et al.}, 
1996), bending waves in Saturn's rings (Gresh
{\it et al.}, 1986; Rosen and Lissauer, 1988), spiral Lin-Shu 
density waves of the type invoked to explain the spiral structure of
disk galaxies, a nonlinear density wave theory, viscous damping, and
a great number of moonlets orbiting inside the optically thick parts 
of Saturn's rings (Colwell, 1994; Spahn {\it et al.}, 1994).
Franklin {\it et al.} (1982) presented an evidence that two previously 
unidentified, yet conspicuous gaps in Saturn's rings lie at distances 
Gaps such as these can be produced in a ring of large bodies or small
uncharged particles by a nonaxisymmetric gravitational field (both of
the above can be associated with the l = m = 3 harmonic), a fact that 
is relevant to models of planetary interiors. 
The F ring is highly stirred by shepherds, and embedded moonlets are
suggested on the basis of charged particle absorption as well as both
azimuthal and radial structure (Cuzzi and Burns, 1988).  Through a
numerical modeling, Kolvoord and Burns (1992)  demonstrated that the 
modest, out-of-plane satellite-induced vertical and horizontal 
distortions of three narrow bands generate a 
structure akin to the ``braided" F ring.\footnote{A systematic, 
uniform search of Voyager 2 
photopolarimeter system data set for 216 significant features of
Saturn's rings with the spatial resolution in the radial direction 
in the ring plane better than 100 m was described by Esposito {\it et
al.} (1987).  Also, Brophy and Rosen (1992) conducted a parallel 
examination of Voyager radio and photopolarimeter occultation 
observations of the Saturn A ring's satellite-excited  density
waves.}  In the later case, both the structure and the origin of
rings can be explained by a multitude of small moons, still unseen, 
orbiting nearby or among the rings (Esposito, 1992).

It is important that the Voyager's
stellar occultation data revealed some {\it indirect} evidence
for structuring in the densest central parts of opaque Saturn's 
B ring down to the $100$ m length scale (Showalter and Nicholson,
1990).  One cannot exclude the existence of such kinds of small-scale
structure in other, low optical depth regions of the system.

Although direct observations on the small scales of interest ($< 100$ m
or so) in Saturn's rings are nonexistent, there is indirect evidence for  
structure on this scale in the A ring.  The reflectivity of the A ring
varies strongly with longitude in Earth-based and Voyager observations
(Cuzzi {\it et al.}, 1984; 
Franklin {\it et al.}, 1987; Dones {\it et al.}, 1993).  The full 
amplitude of this variation peaks at $30-40\%$ in the mid-A ring and
smaller toward the inner and outer ring edges.  By using the International 
Planetary Patrol network, Thompson {\it et al.} (1981)  detected that the
azimuthal brightness variations for the brighter portion of ring A 
increased as the ring tilt decreased from B = 26 deg to less than 16 deg, 
reaching the order of + or 20
detectable for ring B or the outer portion of ring A.  The variation is
generally known as the ``quadrupole azimuthal asymmetry," because the
effect is not symmetric about the ring ansae.  This asymmetry has been
known for several decades.  A simple semiquantitative explanation 
for the quadrupole asymmetry in terms of numerous unresolved density 
``wakes" caused by gravitational interactions of the particles has been
presented by Colombo {\it et al.} (1976) and Franklin and Colombo (1978).
The presence of wakes causes the effective area covered by  particles, 
hence brightness, of the ring to vary at different longitudes (Franklin
{\it et al.}, 1987).  This explanation, which requires some degree of
self-gravitation between nearby orbiting bodies, accounts both for the
presence of the azimuthal brightness variations in Saturn's ring A and
for their absence in ring B.  A bias in the particle distribution and
corresponding photometric effects are thereby produced the latter
corresponding very closely to the variations observed in ring A.
Their absence in ring B is primarily a consequence of the higher 
optical thickness and decreasing importance of self-gravitation in
that ring.

What caused the stratification of the main Saturn's rings?
Various theories have been advanced to explain the creation of 
the small-scale structure in Saturn's rings.  For instance, 
Lin and Bodenheimer (1981), Lukkary (1981), and Ward (1981)
explaned this microstructure by diffusional instabilities 
(or negative diffusion instabilities).
According to Goldreich and Tremaine (1978, 1982), Cuzzi {\it et
al.} (1981), Lissauer (1989), and Goldreich {\it et al.} (1995), 
the fine radial ring structure 
can be associated with resonant forcing by external moons.  
Shu {\it et al.} (1983) developed the
theory of forced bending waves which cause small-scale vertical
corrugations of the local ring plane.  As a matter of fact, a few dozen
spiral density wave trains that are resonantly forced in the plane of
the system under study by external moons have been detected (Smith
{\it et al.}, 1982; Shu {\it et al.}, 1985; Esposito, 1986, 1993).
Lissauer (1985) has found features within Saturn's rings that 
might be produced by vertical resonances of external moons.  However,
it is clear that only a small part of the structure of Saturn's rings
is determined by plane and vertical resonances with satellite orbits.
Also, it has been suggested that a part of the irregular structure
may be due to embedded moonlets orbiting within the ring system
(Lissauer {\it et al.}, 1981; Colwell, 1994; Spahn {\it et al.}, 
1994).  By $N$-body simulations, Osterbart and Willerding (1995)
also concluded that the most promising explanation for the ringlet
structure of the B ring of Saturn is the assumption that a great
number of moonlets within the ring system can trigger trailing
density wakes of the type studied by Julian and Toomre (1966). 
As a matter of fact, a small moon, Saturn's eighteenth satellite has
been discovered embedded within the Encke gap of the ring system
(Showalter, 1991).  But no more satellite has yet been discovered
within the main rings.  In turn, Durisen {\it et al.} (1989) and 
Durisen {\it et al.} (1996) suggested
that some of features of Saturn's rings --- radial optical depth 
structures near the inner edges of Saturn's A and B rings, including
the edges themselves --- can be
produced or maintained by ``ballistic transport," 
that is, radial transport of mass and angular momentum
due to exchanges of ejecta from meteoroid impacts on ring particles. 
But it is clear, these observations and theories do not imply that most 
of the small-scale structure in the rings results from external or
embedded moonlets, diffusional instabilities, ballistic transport,
and other mechanisms may also be proposed.

In regard to the existence of the irregular structure in the Saturnian
ring system, it seems likely that a universal mechanism that will generate
all types of the structure does not exist: there should be several possible
mechanisms.  Moreover, different regions of Saturn's rings may prefer
different types of instabilities of small-amplitude gravity perturbations
for ring generation.

In the following we argue that small-scale, hyperfine $\sim 2 \pi h \stackrel
{<}{\sim} 100$ m structures could be primarily produced by the classical
gravitational Jeans-type instability in low optical depth regions and
a secular dissipative-type instability in high optical depth regions in 
the system under study.\footnote{Generally, the term ``Jeans instabilities"
identifies nonresonant instabilities associated with growing accumulations 
of mass (cf. electrostatic bunching instabilities or a fire-hose instability 
of a plasma).}  Here $h$ is the typical thickness of the system.  Thus, we
propose that the numerous ringlets in the Saturnian ring system are the
manifestation of tightly wound spiral and/or radial density waves in the
disk, which remain quasi-stationary in a frame of reference rotating around
the center of the system at a proper speed.  In this regard, the idea
that the radial gravitational instability of a gas-dust disk may 
have played a vital role in the formation process of the planetary 
system seems to have been first suggested by Ginzburg {\it et al.}
(1972), Polyachenko and Fridman (1972) (see also Fridman and 
Polyachenko (1984, Vol. 2, p. 261)).

As mentioned above, in this paper we will give a self-consistent
asymptotic solution to the kinetic equations of particle dynamics for
a thin, rapidly rotating disk of mutually gravitating identical particles
in Keplerian rotation around a central gravitating mass,
taking into account the effects of interparticle physical collisions.
This work is of general interest for the theory of particulate disks.
Even though some aspects of the results were previously known, the
authors feel that the new technique provides additional physical
insight into the process involved.  The linear stability analysis 
presented in this paper does make predictions about the morphology of
ring structure that could be compared to future Cassini 
measurements --- tightly wound spiral and/or radial wavelike structures 
with size and spacing of the order $2 \pi h$.  This is a first step towards
a general theory of planetary ring dynamics.

\section{Basic equations}

Under the reasonable assumption that the inclusion of motions
normal to the plane makes little difference to the evolution
of the rapidly rotating system of particles, let us consider dynamics of
an infinitesimally thin self-gravitating disk (Griv, 1996; Griv and 
Peter, 1996). This is a valid approximation if one considers perturbations
with a radial wavelength that is much greater than $h/2  \approx c^2
/ \sqrt{4 \pi G \Sigma}$, where $c$ is the mean dispersion of random
velocities, $G$ is Newton's gravitational constant, and $\Sigma$ is the
volume mass density in the mid-plane (Shu, 1970; Ginzburg {\it et al.},
1972; Griv and Yuan, 1997).  Note that it has been shown by 
$N$-body simulations of disk-shaped galaxies of stars that the
inclusion of motions normal to the plane makes little difference to the
evolution of the rapidly rotating thin disk (Hohl, 1978).  (Common dynamical
processes act in the stellar disks of flat galaxies and in Saturn's ring
system of particles; Tremaine (1989)).  This justifies the two-dimensional
treatment of the main part of a rapidly rotating particulate disk.  From
now on, as a simplification, we assume the disk is two-dimensional and,
therefore we consider the special case of waves propagating in the
equatorial plane of the system under study.\footnote{While the theoretical
model is two dimensional, actual self-gravitating disks have finite
transverse dimension.  When the disk is ``strongly magnetized" (rapidly
rotating), motion along the ``magnetic field" (along the vector of the 
angular rotation) may be neglected so that the motion of the particles is 
two dimensional.  In this case the two-dimensional model is appropriate
provided the waves generated are also nearly two dimensional; in general,
in the rapidly rotating disks the coupling between waves propagating in
the plane and along the axis of rotation is expected to be small.  
We suggest that waves in the plane are coupled with waves propagating in 
the normal to the plane direction at the position of the so-called vertical
resonances , and here the self-excitation of spontaneous kinetically-unstable
bending waves is expected (Griv {\it et al.}, 1997a).  Clearly, such effects
can only be studied by using a three-dimensional model.}

The collision motion of an ensemble of identical particles in the plane,
in the frame of reference rotating with angular velocity $\Omega$, can
be described by the Boltzmann equation
(Lin and Shu, 1966; Lin {\it et al.}, 1969; Griv and Peter, 1996)
\begin{eqnarray}
{\partial f \over \partial t} &+& v_r {\partial f \over \partial r} 
+ \left(\Omega + {v_\varphi \over r}\right) 
{\partial f \over \partial \varphi}
+ \left(2 \Omega v_\varphi + { v_\varphi^2 \over r } 
- {\partial \Phi_1 \over \partial r} \right)
{\partial f \over \partial v_r} \nonumber \\
&-& \left({\kappa^2 \over 2 \Omega} v_r + {v_r v_\varphi \over r}
+ {1 \over r}{\partial \Phi_1 \over
\partial \varphi} \right){\partial f \over \partial v_\varphi} 
= \left( {\partial f \over \partial t} \right)_{\mathrm{coll}} ,
\end{eqnarray}
\noindent
where the total azimuthal velocity of the particles is represented
as a sum of $v_\varphi$ and the circular velocity $r \Omega$, 
$\Omega (r)$ is the angular velocity at the distance $r$ from the
planet, and $v_r$ and $v_\varphi$ are the minor residual 
(random) velocities
in the radial and azimuthal directions, respectively, $|v_r|$ and
$|v_\varphi| \ll r\Omega$. Applying the linear theory of disk stability, 
in the kinetic equation (1) the total gravitational potential $\Phi 
(\vec{r}, t)$ is divided into a 
smooth basic part $\Phi_0 (r)$ satisfying the equilibrium condition
$$
\Omega^2 r = {\partial \Phi_0 \over \partial r} ,
$$
\noindent
and a small fluctuating part $\Phi_1 (\vec{r}, t)$ with
$|\Phi_1 / \Phi_0| \ll 1$ for all $\vec{r}$ and $t$. 

The left-hand side of the Boltzmann equation (1) represents the total
time rate of change of the distribution function $f$ along a particle
trajectory in $(\vec{r}, \vec{v})$ space as defined by Langrange's 
system of characteristic equations:
$$
d \vec{r} / d t = \vec{v} \quad \mbox{and} \quad d \vec{v} / d t 
= - \partial \Phi / \partial \vec{r} .
$$

In Eq. (1), $(\partial f / \partial t )_{\mathrm{coll}}$ is the
so-called collision integral which takes into account effects due to the
discrete-point nature of the gravitational charges, or collision effects 
(including diffusion in space and velocity), and defines the change 
of the distribution function $f (\vec{r},\vec{v},t)$ arising from 
ordinary interparticle collisions (in a plasma this term represents the 
change of $f$ arising from collisions with particles at distances
shorter than a Debye length).  The Boltzmann form for the collision
integral is based on an assumption that the duration of a
collision is much less than the time between collisions --- instantaneous  
collisions are considered.  We assume that there are only binary
physical collisions, and momentum is conserved in collisions.  In addition
there is no correlation in motion between the colliding species,
that is, Boltzmann's hypothesis of molecular chaos is adopted. The left-hand
side in Eq. (1) is the total rate of change of the phase space density
following the motion.  The Boltzmann equation then says this density
changes because of the collisions, following the phase space trajectories.

In plasma physics, Lifshitz and Pitaevskii (1981, p. 115) have discussed
phenomena in which interparticle collisions are unimportant, and such a
plasma is said to be collisionless (and in the lowest-order approximation of
the theory one can neglect the collision integral in the kinetic equation).
It was shown that a necessary condition is that $\nu_{\mathrm{c}} 
\ll |\omega|$, where
$\omega$ is the frequency of excited oscillations: then the collision 
operator in the kinetic equation above is small in comparison with $\partial
f / \partial t$.  Lifshitz and Pitaevskii (1981) have pointed out that
collisions may be neglected also if the particle mean free path is large
compared with the wavelength of collective oscillations.  Then the collision
integral in Eq. (1) is small in comparison with the term $\vec{v} \cdot
(\partial f / \partial \vec{r})$.

Equation (1) resembles the Boltzmann equation for a collisional plasma 
in a nonuniform magnetic field, thus the techniques of plasma
theory may be applied.  In plasma physics, methods for
investigating oscillations and the stability of a collisional
system have been developed using either the exact Boltzmann
integral formulation or an approximate collisional term in
the form of Krook (or Bhatnagar {\it et al.}) model.  Reviews of
plasma kinetic theory, taking into account collisions between
particles, are given by Rukhadze and Silin (1969), Mikhailovskii 
(1974), and Alexandrov {\it et al.} (1984).

In general, the Boltzmann equation is nearly intractable
because of the complicated collision integral.  The collisional
term can be approximated in various ways, the simplest is to assume
that it vanishes in the collisionless model.
In this work we use the simple kinetic model when the
exact, but complicated, Boltzmann integral $(\partial f / \partial
t)_{\mathrm{coll}}$ is replaced by an approximate, phenomenological   
term in the form of the Krook model (Shu and Stewart, 1985).
The Krook integral is called a model because it cannot be  
derived from the exact Boltzmann integral but can only be
constructed by general physical reasoning, i.e. the need to satisfy
the laws of conservation of particle number, momentum, and energy.
  
The simple Krook integral in the case of a 
two-dimensional disk of identical particles has the form 
\be
\left( {\partial f \over \partial t} \right)_{\mathrm{coll}}
= - \nu_{\mathrm{c}} (f - f_0) ,
\ee
\noindent
where $f$ is the actual distribution function of particles and 
$f_0$ is the steady-state equilibrium distribution function (Shu 
and Stewart, 1985).  
In Eq. (2), $\nu_{\mathrm{c}}$ plays the role of the
velocity-independent collision frequency; 
$\nu_{\mathrm{c}} = n \langle s v
\rangle$, where $n$ is the number density of particles, $s$ is the
effective radius of a particle, and $\langle \dots \rangle$
denotes the average over particles of all random velocities $v$ in
a Maxwellian distribution.  Results following from Eq. (2) depend on 
$\nu_{\mathrm{c}}$, 
and one must seek other, external means to obtain the value
of $\nu_{\mathrm{c}}$.  In order to conserve particles in this model, we
demand $\int (f - f_0) d \vec{v} d \vec{r} = 0$, since $(f - f_0)$
will eventually decay away.  On the other hand, it will relax the
distribution towards equilibrium with an increase in entropy.
The distribution function $f_0$ is necessarily a Maxwellian-like.
In the case of an infinitesimally thin, differentially rotating ($d \Omega
/ d r \ne 0$) disk with collisions, the function $f_0$ is given by
\be
f_0 = {\sigma_0 (r) \over
2 \pi c_r (r) c_\varphi (r)} \exp \left\{ - {v_r^2 \over
2 c_r^2 (r)} - {v_\varphi^2 \over 2 c_\varphi^2 (r)} \right\} ,
\ee
\noindent
with $c_r$ and $c_\varphi$ being the averaged dispersion of radial 
and azimuthal random velocities of particles, respectively, in
general $c_r \ne c_\varphi$, and $\sigma_0 (r)$ being the equilibrium
surface mass density (Griv, 1996; Griv and Peter, 1996).  In the case
of an almost collisionless system, the function $f_0$ can be constructed
(satisfying the unperturbed part of the kinetic equation)
with the use of the constants $\mbox{I}_1, \mbox{I}_2, \dots$, of the
equilibrium particle motion $f_0 = F_0 (\mbox{I}_1, \mbox{I}_2, \dots)$,
where $F_0$ is, generally speaking, an arbitrary function.  In a system
with frequent collisions, the postcollisional velocities distributed
isotropically, $c_r = c_\varphi$.

The Krook collision integral (2) can be interpreted as the following. 
As a result of scattering, particles are lost at the rate
$\nu_{\mathrm{c}} f$ (an ``absorption" process) and re-emitted at the
rate $\nu_{\mathrm{c}} f_0$ with Maxwellian distribution of the 
postcollisional velocities at the mean surface density.  The
greatest defects of the Krook collision integral is that it cannot
be derived from the Boltzmann integral but can only be constructed
by general physical reasoning and is that in the case of 
small-angle (gravitational or Coulomb) collisions the ``diffusion"
coefficient $\nu_{\mathrm{c}}$ does not fall of with increasing velocity, 
as do
those given by the Fokker-Planck equation (Rosenbluth {\it et al.},
1957; Rukhadze and Silin, 1969).\footnote{The
replacement $\nu_{\mathrm{c}} (\vec{v}) \approx 
\nu_{\mathrm{c}}$ is exact for 
a repulsive
short-range force $\Phi \propto - 1 / r^5$ between the electrically
charged particle and neutral target particles
(Krall and Trivelpiece, 1986, p. 315).}  However,
by replacing the exact Boltzmann collision integral in
the kinetic equation by an approximate term in the form of Krook
model the problem of the stability can be solved with simple methods
similar to those used for the collisionless case: by applying the
usual procedure in particle path integration (Mikhailovskii and
Pogutse, 1966; Mikhailovskii, 1974, Vol. 2; Griv and Chiueh, 1997).

The right-hand side of Eq. (1) as given by Eq. (2) is a gross 
approximation to the collision integrals.  It represents a relaxation
term to a steady state distribution, $f_0$.  The solution will be 
assumed to depend analytically upon $\nu_{\mathrm{c}}$.

It has to be noted that at least in the case of small-angle
collisions the Krook model cannot be used in the
case of perturbations of a particulate disk with too small,  
$k^2 \rho^2 \gg 1$, and too large, $k^2 \rho^2 \ll 1$, wavelengths.
Pitaevskii (1963), Rukhadze and Silin (1969), and Griv {\it et al.} 
(1997b) already explained the problem.  Here $k$
is the wavenumber, $\rho \approx c / \kappa$ is the mean epicyclic
radius (the Larmor radius in a plasma), and
$$
\kappa = 2 \Omega \left( 1 + {r \over 2 \Omega}{d \Omega \over d r}
\right)^{1/2}
$$
\noindent
is the ordinary epicyclic frequency.  Also, the collisonal term (2)
is applicable only in the case of purely elastic collisions, e.g.
in the case of gravitational encounters.  Let us modify Eq. (2) to 
allow the effects of inelastic (physical) collisions.  
Following Shu and Stewart (1985), it is supposed here 
that each collision reduces the magnitude of the
component of the relative velocity along the line of centers of
particles by a factor $\epsilon < 1$, 
where $\epsilon$ is the
coefficient of restitution averaged over all encounters, and is
defined from the equation
\be
3 c_I^2 = (2 + \epsilon^2) c^2 ,
\ee
\noindent
where $c_I$ is the velocity dispersion after collision and
$c$ is the velocity dispersion before collision. A very popular
model of the particles of Saturn's rings is a smooth ice sphere,
whose restitution coefficient $\epsilon$ is quite high, $0.5 <
\epsilon < 1$, $\epsilon$ is decreased as the collision velocity 
increases, and impact velocities are only a few millimetres per 
second (Goldreich and Tremaine, 1982; Bridges {\it et al.}, 1984).
Momentum is conserved in inelastic collisions, so that in the
equation of motion the viscosity can be determined by the 
effective intercollision time $\propto 1 / \nu_{\mathrm{c}}$.
Thus, $c_I < c$, and in Eq. (2) one can replace $f_0$ by a
Maxwellian-like distribution with the postcollisional velocities.

One should keep in mind that without inelastic collisions a planetary
ring cannot come into a state of equilibrium and will be ``heaten up."
Without dissipation the velocity dispersion of random velocities of
colliding particles in a rapidly rotating system grows without limit
and any structures are transient structures only.  The coefficient of
restitution $\epsilon$ in inelastic collisions can be a function of the
(thermal) impact velocity $v$, and this fact should be taken into account 
by modelling in great detail the equilibrium distribution (Goldreich and  
Tremaine, 1978; Shu and Stewart, 1985).  Planetary
rings are gravitationally and collisionally dominated systems, and a
thermal quasi-equilibrium in these systems is achieved via gravitational
(due to gravitational instabilities) ``heating" and viscous (due to
velocity shear) heating and collisional cooling (damping of the impact
velocities due to dissipative collisions).  We discuss this problem in
brief in Sections 4.1 and 4.2.

Shu and Stewart (1985) have estimated the velocity-independent  
frequency of inelastic collisions, using a similar
relaxation-time approximation to the collision term: 
\be
\nu_{\mathrm{c}} \approx {8 \over \pi} \mu \tau ,
\ee
\noindent
where $\mu \approx \sqrt{4 \pi G \Sigma} \approx \Omega$ is the
frequency of vertical epicyclic oscillations and $\tau$ is the normal
optical depth.  In the case of spherical particles with a power-law
distribution of sizes (and masses), $\tau$ should be replaced by
the quantity $\tau_e$ which is smaller than the normal
optical depth by the factor 3 or 4 (Shu and Stewart, 1985).

As one can see, the collision integral (2) will vanish when the
equilibrium Maxwellian distribution is substituted, $f \equiv f_0$.
It is assumed in Eq. (2) that there is no systematic, mean 
movement of particles excepting for the circular rotation. 
The collision integral of the form (2) does not take into account
detailed mechanisms of the inelastic interaction such as spin
degrees of freedom, the particle size distribution, 
and the finite size of the particles (Araki 
and Tremaine, 1986).  Such effects can, of course, be 
included in the analysis if necessary.  Thus, it is not entirely clear
to what extent the results obtained from our study are relevant to the
structure of Saturn's rings.  Nonetheless, we apply our results to the
standard uniform-size hard sphere model (Goldreich and Tremaine, 1978)
for simplicity.

It may be shown by integrating over velocities that the collisional 
term in such a form conserves the number of particles and the momentum
only on the average over a cycle (Griv and Chiueh, 1997).\footnote{  
The more complicated phenomenological Bhatnagar-Gross-Krook  
integral has the form
$$
\left( {\partial f \over \partial t} \right)_{\mathrm{coll}}
= - \nu_{\mathrm{c}} \left( f - {\sigma \over \sigma_0} f_0 \right) ,
$$
\noindent
where $\sigma$ is the local surface density (Bhatnagar {\it et 
al.}, 1954).  In contrast to the simpler case of Krook's collision
integral, the collision term in such a form instantaneously 
conserves the number of particles, the momentum, and the particle's
energy (Griv and Chiueh, 1997).}  Random kinetic energy is always
dissipated on the average over finite times by the inelastic part
of the collisional process (this is because $c_I^2 < c^2$ is 
always true).  The collision term of the form (2) forces the
distribution function to relax in position space upon each collision
to the uniform isotropic Maxwellian distribution 
of postcollisional velocities at the mean density.

Of course, the question that remains is whether the $f_0$ used in the
Krook collisional integral is appropriate to investigate the effects
of frequent and strong collisions; it could probably be strongly 
modified by frequent collisions.  For instance, Wisdom and Tremaine
(1988) show that, as optical depth increases, significant stresses are
communicated through the finite size of closely packed 
particles.\footnote{It was a question posed by the referee of the paper:
whether a simple gas-kinetic theory approach breaks down in the regime
of the high optical depth.}  In a future investigation, several results 
of the kinetic theory of oscillations of a collisional particulate disk
obtained with the aid of the exact Boltzmann integral will be presented 
(Griv, Gedalin, and Yuan, 1999, in preparation).  
This will show clearly which of the results obtained with the
Krook model collision integral are qualitatively correct.

Perturbations in the gravitational field cause perturbations
to the particle distribution function.  In the linear approximation,
one can therefore write
$f(\vec{r}, \vec{v}, t) = f_0 (r, \vec{v}) + f_1(\vec{r}, \vec{v}, t)$,
where $|f_1| \ll f_0$ and $f_1$ 
is a function rapidly oscillating in space and time.  If an initial
perturbation grows, the system is called unstable.  The function $f_0$
describes the differentially rotating ``background" against which   
small perturbations develop.  Initially, the disk is in an equilibrium,
$\partial f_0 / \partial t = 0$.

The linearized kinetic equation (1) for the perturbed distribution
function $f_1 (\vec{r}, \vec{v}, t)$ becomes
\be
{d f_1 \over d t} = {\partial \Phi_1 \over \partial r} {\partial f_0
\over \partial v_r} + {1 \over r}{\partial \Phi_1 \over \partial 
\varphi} {\partial f_0 \over \partial v_{\varphi}} - 
\nu_{\mathrm{c}} f_1 ,
\ee
\noindent
where $d / d t$ is taken along the unperturbed orbits of particles 
in the local rotating frame.  In turn, the linearized kinetic 
equation (1) for the unperturbed distribution function
$f_0 (r, \vec{v})$ takes the form
$$ 
v_r {\partial f_0 \over \partial r} + \left( 2 \Omega v_\varphi +
{v_\varphi^2 \over r} \right) {\partial f_0 \over \partial v_r}
- \left( {\kappa^2 \over 2 \Omega} v_r + {v_r v_\varphi \over r}
\right) {\partial f_0 \over \partial v_\varphi} = 0 .
$$ 
\noindent
Initially the disk is in equilbrium, $\partial f_0/\partial t=0$.  

Generally, the wave is not plane.  However, if we assume that the
solution is nearly a plane wave, the expression for the field may
be written in the form
$$
\aleph = \delta \aleph (r) e^{i {\cal{A}} (r) } 
$$
\noindent
(we omit the $\Re$; it is understood that the real part of all
expressions is taken).  In homogeneous media ${\cal{A}} (r) = 
k_r r$.  If the medium is sufficiently slowly varying, the
simple plane wave solution above represents a good starting point.

If geometrical optics (or the standard WKBJ method in quantum mechanis)
is applicable, the amplitude $\delta \aleph$ is, generally speaking, a
function of the coordinates and time, and the phase ${\cal{A}}$,
which is called the eikonal, is a large quantity.  Following the
WKBJ method, the perturbations will be taken to be of the form
$$
f_1, \Phi_1 \sim \sum_{m=-\infty}^\infty
\delta f_m (r), \delta \Phi_m (r), \exp \{ -i \omega
t + i m \varphi + i k_r r \} , 
$$
\noindent
$|k_r| r \gg 1$, which corresponds
to the fact that in each small region of space (and each small
interval of time) the wave can be considered as plane. 
Evidently $f_1$ and $\Phi_1$ are periodic functions of
$\varphi$, and hence $m$ must be an integer.  
Consideration will be limited to the region between the turning 
points in a disk (the transparency region).  Here $\omega$ is the
frequency of excited waves, $m$ is the positive azimuthal 
mode number, which gives the number of spiral arms (for
axisymmetric perturbations $m = 0$), and $k_r$ and $k_\varphi
\equiv m / r$ are the radial and azimuthal components of the
wavevector $\vec{k}$.  In the framework of the linear theory
we are interesting, we can select
one of the Fourier harmonics: $\delta f, \delta \Phi 
\exp \left( i k_r r + i m \varphi - i \omega t \right)$.
In the local approximation of the WKBJ
method (with no derivatives of $\delta \aleph$ and neglecting
$d^2 {\cal{A}} / d r^2$, e.g. Swanson (1989, p. 13)), 
$\delta f$ and $\delta \Phi$ are constants.  In addition,
perturbations with a wavelength $\lambda = 2 \pi / k_r$
such that $h \stackrel{<}{\sim} \lambda \ll R$ are investigated, 
where $R$ is the radial size of of a system.  Then, the disk
may be regarded as infinitesimally thin.
	
In the case of the two-dimensional disk we are interested in,
the Poisson equation is
$$
{\partial^2 \Phi \over \partial r^2} + {1 \over r}{\partial \Phi
\over \partial r} + {1 \over r^2}{\partial^2 \Phi \over \partial
\varphi^2} + {\partial^2 \Phi \over \partial z^2} = 4 \pi G
\sigma \delta (z) ,
$$
\noindent
where $\sigma (r)$ is the
surface mass density, and $\delta (z)$ is the Dirac delta-function.  
The improved asymptotic, $k_r^2 \gg k_\varphi^2$, 
Lin-Shu type solution of this equation may be written in the form
(Lau and Bertin, 1978; Lin and Lau, 1979; Bertin, 1980)
\be
\sigma_1 (r) = - \mbox{sign} (k) {k \Phi_1 (r) \over 2 \pi G} 
\left\{ 1 - {i \over k_r r} {d \ln \over d \ln r} \left[ r^{1/2}
\delta \Phi \right] \right\} ,
\ee
\noindent
where $\sigma_1$ is the 
small perturbation of the equilibrium surface density.  The wave 
vector $\vec{k}$ in the two-dimensional system is given by 
$k^2 = k_r^2 + k_\varphi^2$ and the angle
between the direction of the wave front and the tangent to the
circular orbit of the particle (the pitch angle) is defined by
$$
\tan \psi = {k_\varphi \over k_r} = {m \over r k_r} .
$$
\noindent
Solution (7) which determines the perturbed
surface mass density required to support the perturbed
gravitational potential was found in the asymptotic Lin-Shu type
(Lin and Shu, 1966; Shu, 1970) approximation of moderately tightly
wound perturbations $\tan^2 \psi \ll 1$ which has traditionally
been used in all such investigations (see Bertin and Mark (1978),
Lin and Lau (1979), and Bertin (1980) for a discussion).
The improved solution (7) of the Poisson equation, which includes
effects of finite inclination of spiral arms ($k_\varphi \ne 0$),
is accurate to two orders in the familiar Lin-Shu asymptotic 
approximation.

In order to find the perturbed distribution $f_1$, it is convenient
to integrate Eq. (6) along the unperturbed trajectories of particles.
Since this is difficult to do exactly, epicyclic orbits and series
expansions are used instead.  Defining the epicycle phase $\phi_0$ 
at $t=0$ by $[v_r , v_{\varphi} ] = [v_{\perp} \cos \phi_0 ,
~(\kappa / 2 \Omega) v_\perp \sin \phi_0]$, where $\phi_0$ and
$v_\perp$ are constants of integration, the solution for the
ordinary Lindblad elliptic-epicyclic trajectories in an 
unperturbed central force field of Saturn can be expressed as
\begin{eqnarray}
r &=& r_0  
- {v_\perp \over \kappa} \left[ \sin(\phi_0 - \kappa t)  
- \sin \phi_0 \right] , \nonumber \\
\varphi &=& \Omega t + {2 \Omega \over \kappa}  
{v_\perp \over \kappa r_0} \left[ \cos(\phi_0 - \kappa t)  
- \cos \phi_0 \right] ,
\end{eqnarray}
\noindent
where $r_0$ is the radius of the chosen circular orbit in the $(r,
\varphi)$ plane.  The equations above describe the small departure
of the actual radius $r (t)$ from $r_0$,
which is chosen so that the constant of areas for the circular orbit
$r_0^2 \dot{\varphi}_0$ (and $r_0 \dot{\varphi}_0^2 = \left( \partial 
\Phi_0 / \partial r \right)_0$) is equal to the angular momentum 
integral $r^2 \dot{\varphi} = \mbox{const}$.
In Eqs. (8), $v_\perp / r_0 \kappa \sim \rho / r_0 \ll 1$, and
$\rho \simeq v_\perp /\kappa$ is the epicyclic radius (Grivnev,
1988; Griv and Peter, 1996).  The zeroth order approximation is simply
a circle on which the particle moves with angular velocity $\Omega =
V / r$ and rotational velocity $V (r) = (r d \Phi_0 / d r)^{1/2}$.
The first-order Lindblad
epicyclic theory superposes on this regular rotational motion
harmonic oscillations both in the radial and azimuthal directions with
a characteristic frequency $\kappa$ called the epicyclic frequency,
Eqs. (8)
(cf. a simple cyclotron gyration of a charged particle of a plasma).
This describes a circular orbit of a guiding center.\footnote{
The postepicyclic orbits take into account the second-order effects
of the orbital eccentricity and describe additional directional
motions of particles, i.e. oscillations with combined epicyclic
frequencies, the small displacement of the center of the epicycle
along the field gradient, and the small non-oscillatory drifting
motions both along the circular orbit and along the epicycle (Griv,
1996; Griv and Peter, 1996; Griv {\it et al.}, 1999a).} 
In the Saturnian ring disk $\kappa \approx \mu \approx \Omega$.

The closure condition for the orbit up to first order in Lindblad's
epicyclic theory may be written as
$$
{\kappa \over \Omega} = {n \over s} ,
$$
\noindent
where $n, s$ are positive integers.  Obviously, the above condition
determines only two types of fields in which all orbits are closed:
the limit of Keplerian rotation, $n=s=1$, and the limit of rigid rotation,
$n=2s=2$.  In the former limit, after one complete revolution along the
circular orbit and one complete revolution along the epicycle, the
particle will occupy its original position.  

From Eqs. (8) it easy to find $u_r$ and $u_\varphi$, the 
components of the particle's random velocity relative to the 
planetary center.  In the lowest order, the solutions are
$$
u_r = v_\perp \cos (\phi_0 - \kappa t) , \quad u_\varphi = 
{2 \Omega \over \kappa} v_\perp \sin (\phi_0 - \kappa t) .
$$

To integrate Eq. (6) over $t$, we need to determine the
components $v_r$ and $v_\varphi$ of the particle's 
velocity at each point relative to the local standard of rest.
Since the circular velocity in this rotating system, is $- r_1 r_0
(d \Omega / d r)$, where $r_1 = r - r_0$ (Spitzer and Schwarzschild,
1953; Binney and Tremaine, 
1987, p. 120; Griv and Peter, 1996), we have
\be
v_r = u_r ; \quad v_\varphi = u_\varphi + r_0 {v_\perp \over 
\kappa} {d \Omega \over dr} \sin (\phi_0 - \kappa t) \simeq
{\kappa \over 2 \Omega} v_\perp \sin (\phi_0 - \kappa t) .
\ee
  
Additionally, to perform the integral in Eq. (6) an expression for 
the equilibrium distribution function $f_0$, Eq. (3), is needed.  For  
an infinitesimally thin disk with rare collisions between particles,
$\nu_{\mathrm{c}} \ll \Omega$, we choose the Schwarzschild 
distribution function
(the anisotropic Maxwellian distribution function) satisfying
the unperturbed part of the kinetic equation (Shu, 1970)
$$
f_0 = {2 \Omega (r_0) \over \kappa (r_0)} {\sigma_0 (r_0) 
\over 2 \pi c_r^2 (r_0)}
\exp \left\{ - {v_\perp^2 \over 2 c_r^2 (r_0)} \right\} .
$$
\noindent
The function $f_0$ is a function of the epicyclic constants of 
motion ${\cal{E}} = v_\perp^2 / 2$ and $r_0$, and the quantity
$r_0$ represents approximately the $r$-coordinate of the particle
guiding center.  For an infinitesimally thin disk with frequent
collisions between particles, 
$\nu_{\mathrm{c}} \gg \Omega$, we choose the
isotropic Maxwellian distribution with the postcollisional velocities
\begin{equation}
f_0 = {\sigma_0 (r_0) \over 2 \pi c^2 (r_0)} \exp \left\{ 
- {v^2 \over 2 c^2 (r_0)} \right\} .
\end{equation}
\noindent
In the first equation of the system (10) the fact was used that the
radial and azimuthal dispersions in Eq. (3) are not independent but,
according to Lindblad's theory of epicyclic orbits (Eqs. [8] and 
[9]), are related in the rotating frame through
$$
c_r = (2 \Omega / \kappa) c_\varphi .
$$
\noindent
Of course, this relation is valid only when $\nu_{\mathrm{c}}
\ll \kappa$ or $\Omega$.  In the opposite limit of frequent
collisions (the hydrodynamical limit), when $\nu_{\mathrm{c}}
\gg \kappa$ or $\Omega$,
$$
c_r = c_\varphi .
$$

Hence, in the case of rapidly rotating disk with rare interparticle
collisions the partial  derivatives in Eq. (6) transform as follows 
(see also Shu (1970)):
$$
{\partial \over \partial v_r} = v_r {\partial \over \partial 
{\cal{E}} } ; \quad
{\partial \over \partial v_\varphi} \approx \left(
{2 \Omega \over \kappa} \right)^2 v_\varphi {\partial \over 
\partial {\cal{E}} } + {2 \Omega \over \kappa^2} {\partial 
\over \partial r} .
$$
\noindent 
Using above equations, the solution of the
kinetic equation (6) in the considered case of a spatially
inhomogeneous, $\partial f_0 / \partial r \ne 0$, almost  
collisionless disk may be written in the form (Griv, 1996;
Griv and Peter, 1996; Griv and Chiueh, 1997)
\be
f_1 = e^{-\nu_{\mathrm{c}} t} \int_{-\infty}^t d t^\prime 
e^{\nu_{\mathrm{c}} t^\prime} \left( \vec{v}_\perp {\partial \Phi_1
\over \partial \vec{r}} {\partial f_0 \over \partial \cal{E}} 
+ {2 \Omega \over \kappa^2}
{1 \over r_0} {\partial \Phi_1 \over \partial \varphi}
{\partial f_0 \over \partial r} \right) .
\ee

\section{The generalized dispersion relation}

Now it is possible to integrate Eq. (11) along the unperturbed
trajectories (8), $\vec{R} (\vec{r}^\prime ,\vec{v}^\prime,t)$,
that end at $\vec{R} (\vec{r}, \vec{v}, t)$ when $t^\prime
\rightarrow t$, using the equlibrium distribution functions
and relation
$$
J_{l-1} (\chi) + J_{l+1} (\chi) = {2 l \over \chi} J_l (\chi) ,
$$
\noindent
where $J_l (\chi)$ is the Bessel function of the first kind of 
the order $l$.  The method of integration has been described,
e.g. by Krall and Trivelpiece (1987, p. 395) and Swanson (1989, 
p. 142).  The following expression may be easily obtained:  
\be
f_1 = - \Phi_1 (r_0) \left[ \kappa {\partial f_0 \over \partial 
\cal{E}} \sum_{l=-\infty}^\infty l {J_l^2
(k_* v_\perp / \kappa) \over \omega_* - l \kappa + 
i \nu_{\mathrm{c}}} 
+ {2 \Omega \over \kappa^2} {m \over r_0}
{\partial f_0 \over \partial r}
\sum_{l=-\infty}^\infty 
{J_l^2 (k_* v_\perp / \kappa) \over \omega_*
- l \kappa + i \nu_{\mathrm{c}}} \right] ,
\ee
\noindent
where
$$
k_* = k \left\{ 1 + \left[ \left( 2 \Omega 
/ \kappa \right)^2 - 1 \right] \sin^2 \psi \right\}^{1/2}
$$
\noindent
is the effective wavenumber and $\omega_* = \omega - m \Omega$ is
the Doppler-shifted frequency of excited waves as seen by a particle
in a rotating frame.  The addition of the Krook term is equivalent
to the change 
$$
\omega_* \rightarrow \omega_* \left( 1 + 
i {\nu_{\mathrm{c}} \over \omega_*} \right)
\quad \mbox{and} \quad
\left| {\nu_{\mathrm{c}} \over \omega_*} \right| \ll 1.
$$
\noindent
Clearly, 
if $|\nu_{\mathrm{c}} / \omega_*|$ is small enough and $|\omega_*|
\stackrel{<}{\sim} \Omega$ we can ignore these collisions.  Then,
the replacement $\omega_* \rightarrow \omega_* + 
i \nu_{\mathrm{c}}$ would
merely introduce some collisional damping with the modes.

We assume that the perturbations vanish as
$t^\prime \rightarrow - \infty$ and take some values at $t$, 
so we may neglect the effects of the initial conditions.
The method of integration has also been described
by Griv (1996), Griv and Peter (1996), and Griv and Chiueh (1997). 
It is convenient to write the eigenfrequency in a form
of the sum of the real part $\Re \omega_*$ and the
imaginary part $i \Im \omega_*$.  In accordance with the
definition of perturbations $f_1$, $\Phi_1$, and $\sigma_1$, 
the existence of solutions with $\Im \omega_*$ greater than zero 
implies instability; the solutions with $\Im \omega_* = 0$ describe
long-lived natural oscillations.  The solutions with
$(\Re \omega_*)^2 < 0$ describe the aperiodic instabilities. 
Integrating Eq. (12) over velocity space 
$$
\int_{-\infty}^{\infty} d v_r \int_{-\infty}^{\infty}
f_1 d v_\varphi = 2 \pi {\kappa \over 2 \Omega}
\int_0^{\infty} f_1 v_\perp d v_\perp \equiv \sigma_1
$$
\noindent
and equating the result to the ``in-phase" perturbed surface 
density given by the asymptotic solution of the Poisson equation 
(7) $\sigma_1 = - |k| \Phi_1 / 2 \pi G$, the generalized Lin-Shu
type dispersion relation $\omega_* = \omega_* (k_*)$, is obtained
\be
{k^2 c_r^2 \over 2 \pi G \sigma_0 |k|} =  - \kappa \sum_{l = -
\infty}^{\infty} l {e^{-x} I_l (x) \over \omega_*
- l \kappa + i \nu_{\mathrm{c}}} 
+ 2 \Omega {m \rho^2 \over r_0 L}
\sum_{l=-\infty}^\infty {e^{-x} I_l (x) \over \omega_*
- l \kappa + i \nu_{\mathrm{c}}} ,
\ee
\noindent
where $x = k_*^2 c_r^2 / \kappa^2 \simeq k_*^2 \rho^2$, 
$I_l (x)$ is the modified Bessel function of the order $l$, and
$$
\left| L \right| \approx \left|
\partial \ln \left( 2 \Omega \sigma_0 / \kappa c_r^2 \right)
/ \partial r \right|^{-1}
$$
is the radial scale of a spatial inhomogeneity.  In the local 
version of the WKBJ method, we are using, $|k_r|^{-1} < |L| < r$.
In the derivation of Eq. (13), the following formula has been
used
$$
\int_0^\infty e^{- r^2 x^2} J_l (\alpha x) J_l (\beta x) =
{1 \over 2 r^2} \exp \left( - {\alpha^2 + \beta^2 \over 4 r^2}
\right) I_l \left( {\alpha \beta \over 2 r^2} \right) .
$$
\noindent
Lynden-Bell and Kalnajs (1972) first derived the generalized
Lin-Shu type dispersion relation (13) for the case of
non-axisymmetric Jeans-type perturbations by neglecting the
effect of collisions, $\nu_{\mathrm{c}} = 0$, and inhomogeneity, $L  
\rightarrow \infty$ (Eq. [A11] in their paper).

We pay attention mainly to the long-wavelength oscillations,
$x \equiv k_*^2 \rho^2 \le 1$, the case of epicyclic radius 
small compared with wavelength
(but, of course, in order to be appropriate for WKBJ wave
we consider the perturbations with $|k| r \gg 1$).  In this limit,
one can use the following asymptotic expansion of the modified
Bessel functions $I_0 (x) \approx 1 + x^2 / 4$, $I_1 (x) \approx
x / 2$, and $\exp (- x) \approx 1 - x + x^2 / 2$.  The
short-wavelength perturbations, $k_*^2 \rho^2 \gg 1$, are not
as dangerous in the problem of disk stability
as oscillations with $k_*^2 \rho^2 \le 1$, since they lead only
to very small-scale perturbations of the density with the
radial-scale $\lambda_r \ll 2 \pi \rho$.  For the
parameters of Saturn's rings the velocity dispersion of the
largest ring particles is $\stackrel{<}{\sim} 0.5$ cm$\,$s$^{-1}$
and the mean epicyclic radius is $\rho \stackrel{<}{\sim} 10$ m, 
thus, $\rho$ is of the order the largest particle size 
(Esposito, 1986, 1993).  It makes
little sense to speak of collective effects on scales smaller than 
the finite-sized particles.  
The asymptotic expansion of the Bessel functions
$I_l (x)$ in the short-wavelength limit, $x = k_*^2 c_r^2 / \kappa^2
\approx k_*^2 \rho^2 \gg 1$, the case of epicyclic radius that is large
compared with wavelength:
$$ 
I_l (x) \simeq {e^x \over \sqrt{2 \pi x} } \left[ 1 + O
\left( {1 \over x} \right) \right] ,
$$
\noindent
while in a more rigorous approximation $I_l (x)$ is a monotonically
decreasing function of $l$ for a fixed $x \gg 1$.

In Eq. (13),
the functions $\Lambda_l (x) = e^{-x} I_l (x)$ appear commonly in
a theoretical treatment of Maxwellian plasmas in a magnetic field.
It is instructive to note: (i) $0 \leq \Lambda_l (x) \leq 1$;  (ii)
$\Lambda_0 (x)$ decreases monotonically from $\Lambda_0 (0) = 1$;
and (ii) $\Lambda_l (x)$ for $l \ne 0$ starts from $\Lambda_l (0)
= 0$, reaches a maximum, and then decreases.


The dispersion relation (13) describes the ordered behavior 
of a medium near its metaequilibrium state and generalizes the
standard Lin-Shu dispersion relation 
(Lin and Shu, 1966; Lin {\it et al.}, 1969; Shu, 1970) for
nonaxisymmetric perturbations, $\sin \psi \ne 0$, to the case where 
physical collisions occur in an inhomogeneous disk.  
The effects of tangential gravitational 
forces (pitch angle dependent effects, $\psi \ne 0$) 
for the collisionless model of a 
galactic disk have previously been analysed by Lau and
Bertin (1978) and Lin and Lau (1979) in the framework of the
hydrodynamical approach and  Bertin and Mark (1978),
Bertin (1980), Morozov (1980, 1981),
Griv (1996), and Griv and Peter (1996) in stellar dynamics.  (As
mentioned above, in galaxies and Saturn's rings $c_r / r_0 \Omega
\ll 1$, and therefore the systems may be treated as an almost cold
gas.  Thus with the exceptions of resonant regions a kinetic
description yields results no different from those obtained
hydrodynamically.)
In sharp contrast to the original Lin-Shu dispersion relation (Lin
and Shu, 1966; Lin {\it et al.}, 1969; Shu, 1970), Eq. (13) is valid
for relatively open spiral waves
$(k_r \stackrel{>}{\sim} k_\varphi)$, and in addition
describes the effects of interparticle collisions in a spatially
inhomogeneous medium.

The dispersion relation (13) is complicated: the basic dispersion 
relation above is highly nonlinear in the frequency $\omega_*$.  
To see the physical meaning of solutions of Eq. (13), one does not
need the exact solutions.  Rather, in order to deal with the most
interesting oscillation types, let us consider
various limiting cases of perturbations described by some simplified
variations of Eq. (13).  For instance, similar to the plasma physics  
method, it is sufficient to consider only the principal part
of the disk between the inner $l=-1$ and outer $l=1$ Lindblad
resonances (Shu, 1970; Morozov, 1980; Griv {\it et al.}, 1999a); 
the treatment of the Lindblad resonances as well as the 
corotation resonance is beyond the scope of the present paper (Griv,
1996).\footnote{Resonances are places where linearized equations describing
the motion of particles do not apply.  In the vicinity of the resonances it 
is necessary to use nonlinear equations, or to include terms of higher orders
into the approximate form of the equations.  The former approach was
adopted by Contopoulos (1979) and the latter one was adopted by Griv
(1996), Griv and Peter (1996), Griv {\it et al.} (1997a), and Griv
{\it et al.} (1999a).}  In addition, we consider the case of weakly
inhomogeneous medium, when the second term on the right-hand side in
Eq. (13) is only a small correction.

\section{The Jeans-type and dissipative-type oscillations}

Two different cases may be considered in Eq. (13): (a) weak
$\omega_*^2 \gg 
\nu_{\mathrm{c}}^2$ and rare 
$\nu_{\mathrm{c}}^2 \ll \Omega^2$, and (b)
strong $\omega_*^2 \ll 
\nu_{\mathrm{c}}^2$ and frequent $\nu_{\mathrm{c}}^2 \gg \Omega^2$
collisions.  When $\omega_*^2
\gg \nu_{\mathrm{c}}^2$ and 
$\nu_{\mathrm{c}}^2 \ll \Omega^2$, the collisions cause
only small corrections of the perturbed distribution function and
in the zero-order aproximation of the theory all dissipative
effects may be ignored.  This is just the opposite of the procedure in 
ordinary gas dynamics, where collisions are the dominant effect.  This
approach is valid for high ``temperatures" (the random velocity  
temperatures) and low densities, when the
mean potential between neigboring particles is small compared with the
thermal energy.  Collisions are to be relatively unimportant and hence
the form of the collision integral can be grossly approximated. 
In the opposite limiting case of strong and frequent  
collisions, the corrections are large and 
dissipative effects play the main role in a system dynamics.

\subsection{The Jeans oscillations --- weak and rare collisions 
($\omega_*^2 \gg \nu_{\mathrm{c}}^2$ and 
$\nu_{\mathrm{c}}^2 \ll \Omega^2$)}

Such low optical depth regions can be found in the C ring at
distances $r < 92 \, 000$ km from Saturn's center, where the average
optical depth $\tau < 0.1$, in the inner portions of the densest
B ring at distances $r < 100 \, 000$ km,
where $\tau = 0.5-0.8$, and 
in the A ring (including the Encke gap) at distances $ r > 122
\, 000$ km, where $\tau \le 0.5$ (Esposito, 1986, Fig. 2 in his paper).

The equation (13) can be represented in the simplest 
form ($|l| \le 1$)
\be
(\omega_* + i \nu_{\mathrm{c}})^3 - (\omega_* + i \nu_{\mathrm{c}})
\omega_{\mathrm{J}}^2 + \omega_{\mathrm{gr}} \kappa^2 = 0 ,
\ee
\noindent
where the square of the Jeans frequency $\omega_{\mathrm{J}}$ is 
\be
\omega_{\mathrm{J}}^2 \approx \kappa^2 - 2 \pi G \sigma_0 |k| F (x) . 
\ee
\noindent
In Eq. (15), $F (x) = (2 \kappa^2 / k^2 c_r^2) \exp ( - x)
I_1 (x)$ is the so-called ``reduction factor," and $F (x) \to 1$
in a dynamically cold system ($c_r = 0$) and decreases with increasing
$x$ (increasing the velocity dispersion) in a dynamically hot disk
($c_r > 0$).  The reduction factor takes into account the fact that the
wave field affects only weakly the particles with high random velocities.
On the left-hand side in Eq. (14), $|\omega_*|$ and 
$|\omega_{\mathrm{J}}| \gg 
\nu_{\mathrm{c}}$.

Also in Eq. (14),
$$
\omega_{\mathrm{gr}} = 2 \Omega e^{-x} I_0 (x) {2 \pi G \sigma_0
|k| \over k^2 c_r^2} {m \rho^2 \over r_0 L}
$$
\noindent
is the frequency of the so-called gradient oscillations.

In general, Eq. (14) describes two ordinary Jeans branches of
oscillations --- the most important long-wavelength branch, $x = k_*^2
c_r^2 / \kappa^2 \stackrel{<}{\sim} 1$, and the short-wavelength one,
$x > 1$ --- and a new gradient branch of oscillations modified by 
collisions (Griv and Chiueh, 1997). 
The Jeans instability occurs when $\omega_{\mathrm{J}}^2 < 0$. 
In the current subsection, we study the 
physics of this instability and the condition of instability
taking into account the additional effects of rare and weak 
collisions and inhomogeneity.  Note that in plasma physics
an instability of the Jeans type is known as the negative-mass
instability of a relativistic charged particle ring or the diocotron
instability of a nonrelativistic ring that caused azimuthal clumping 
of beams in synchrotrons, betatrons, and mirror machines (Davidson,
1992).

Equation (14) is cubic in $\omega_* + i \nu_{\mathrm{c}}$, but a
useful limit is $\kappa^2 \stackrel{>}{\sim} 
\omega_{\mathrm{J}}^2 \gg \omega_
{\mathrm{gr}}^2$.
Finally, analyzing the simplified dispersion relation (14),
it is useful to distinguish between the cases of axisymmetric
($m=0$) and nonaxisymmetric ($m \ne 0$) perturbations.

From relation (14) in the frequency range 
\be
|\omega_*^3| \sim 
|\omega_{\mathrm{J}}^3| \gg \left| \omega_{\mathrm{gr}}
\right| \kappa^2 
\quad \mbox{and} \quad |\omega_*| \gg \nu_{\mathrm{c}} 
\ee
\noindent
the dispersion law for the Jeans branch of oscillations is
\be
\omega_{* 1,2} \approx \pm p 
|\omega_{\mathrm{J}}| - \omega_{\mathrm{gr}}
{\kappa^2 \over 2 \omega_{\mathrm{J}}^2} - i \nu_{\mathrm{c}} ,
\ee
\noindent
where $p=1$ for Jeans-stable perturbations with 
$\omega_{\mathrm{J}}^2 > 0$ 
and $p=i$ for Jeans-unstable perturbations with 
$\omega_{\mathrm{J}}^2 < 0$.  
In Eq. (17), the term involving
$\omega_{\mathrm{gr}}$ is the small correction, and
in general $\omega_{\mathrm{J}}^2 \sim \kappa^2 / 2$.
To repeat ourselves, in this subsection only
oscillations in a disk with rare collisions are considered.  From
Eq. (17) one concludes that both Jeans-stable 
($\omega_{\mathrm{J}}^2 > 0$)
and Jeans-unstable 
($\omega_{\mathrm{J}}^2 < 0$) perturbations will weakly decay 
as a result of rare collisions.  Accordingly, a spatial inhomogeneity
will not influence the stability condition of Jeans modes.
Because oscillations
in the range (16) are being considered here, the collisional
correction $\sim \nu_{\mathrm{c}}$ and the inhomogeneity correction 
$\sim \omega_{\mathrm{gr}} \sim L^{-1}$
to the Jeans frequency $|\omega_J | \stackrel{<}{\sim} \Omega$
are small, $\sim \nu_{\mathrm{c}} \ll \Omega$.

In the weakly inhomogeneous disk considered in the paper, from Eq. (17)
it follows that the Jeans-unstable perturbations grow in an oscillatory
way, $\Re \omega_{* 1,2} \ne 0$.  The gradient of macroscopic parameters
$\Omega$, $\sigma_0$, $c_r$, and azimuthal mode number $m$ determine the
small real part of such Jeans-unstable modes, $|\Re \omega_{* 1,2} /
\Im \omega_{* 1,2}| \ll 1$.

Thus, it is found that weak and rare collisions between particles lead
to the damping of Jeans modes in a particulate disk.  The effect
is not great: the time necessary 
for the perturbation amplitude to fall to $1/e$ of
its initial value is about the collision time, $\nu_{\mathrm{c}}^{-1}$.
This is much longer than, for instance, the
characteristic time of a single revolution in Saturn's rings,
$\sim \Omega^{-1}$.  The Jeans instabilities are fastest in the
weakly collisional regime, in the sense that their growth rates
are slowed down for increasing interparticle collisions.

It follows from Eq. (17) that the collisional effects do not depend 
on the wavenumber $k$.  The latter contradicts our recent results 
obtained with the exact Landau integral of collisions (Griv {\it et al.},
1997b): in fact, the collision frequency $\nu_{\mathrm{c}}$ used here should be
replaced in the case of small-angle collisions by the effective collision 
frequency $\nu_{\mathrm{eff}} \approx \nu_{\mathrm{c}} k_*^2 \rho^2$,
which describes properly the more rapid collisional smoothing of the
small-scale inhomogeneities of the particle distribution function,
$k_*^2 \rho^2 \rightarrow \infty$, and $\nu_{\mathrm{eff}} 
\rightarrow 0$ for long-wavelength perturbations, $k_*^2 \rho^2
\rightarrow 0$.  Therefore, interparticle collisions are poorly
represented by an approximate method presented here.  The results
obtained in this subsection indicate only a tendency of Jeans
perturbations to be damped in a collisional system, and the damping
rate given by Eq. (17) is correct only to the order of magnitude. 

The Jeans-type perturbations can be stabilized by the random velocity
dispersion.  Indeed, if one recalls that such unstable perturbations
are possible only when $\omega_{* 1,2}^2 \simeq \omega_J^2 < 0$, then
by using the condition $\omega_{* 1,2}^2 \ge 0$ for all
possible $k$, a local stability criterion against arbitrary
Jeans-type perturbations can be written in the form (Morozov, 1980,
1981; Griv, 1996; Griv and Peter, 1996)
\be
c_r \ge  c_{\mathrm{\small T}} 
\left\{ 1 + \left[ \left( 2 \Omega / \kappa 
\right)^2 - 1 \right] \sin^2 \psi \right\}^{1/2} 
\approx c_{\mathrm{\small T}} \left[ 1 + 3 \sin^2 \psi \right]^{1/2} ,
\ee
\noindent
where $c_{\mathrm{\small T}} = 3.36 G \sigma_0 /\kappa$  is the
well-known Toomre's (1964) critical velocity dispersion to suppress
the instability of only axisymmetric (radial) 
gravity  perturbations, and the fact is used that in
Saturn's rings $2 \Omega / \kappa \simeq 2$.  The stability
criterion thus obtained represents generalization of Toomre's
(1964) criterion to the case when additionally nonaxisymmetric
(spiral) gravity perturbations ($\psi \ne 0$) are taken into
account.\footnote{To obtain Eq. (18), one first finds the
critical wavenumber $k_{\mathrm{crit}} \approx (\kappa / 2 
\Omega)(1 / \rho)$ from the relation $\partial
\omega_J^2 / \partial k = 0$, corresponding to the minimum on 
the dispersion curve (15).  Then this $k_{\mathrm{crit}}$
is substituted into
the dispersion relation and from the condition $\omega_J^2 \ge 0$
the critical velocity dispersion is found.  This critical velocity
dispersion will stabilize arbitrary but not only axisymmetric Jeans
perturbations.}  The parameter $\{ 1 + [ ( 2 \Omega / \kappa )^2 -
1 ] \sin^2 \psi \}^{1/2}$ is an additional stability parameter that
depends on both the pitch angle $\psi$ and the amount of differential
rotation in the disk $d \Omega / d r$ (cf. the parameter ${\cal{J}}$
introduced by Lau and Bertin (1978), Lin and Lau (1979), 
and Bertin (1980)). 

As one can see from Eq. (18), the modified critical velocity
dispersion $c_{\mathrm{crit}}$ grows with $\psi$.  Consequently, in
order to suppress the most ``dangerous," in the sense of the loss of
gravitational stability, very open nonaxisymmetric perturbations with
$\psi > 45^\circ$, $c_{\mathrm{crit}}$ should obey the following
generalized local stability criterion:
$$
c_r \ge c_{\mathrm{crit}} = {2 \Omega \over \kappa} c_{\mathrm{T}} .
$$
\noindent
In Saturn's rings $2 \Omega / \kappa \approx 2$.
A relationship exists between Eq. (18) and what
Toomre (Toomre, 1977, 1981; Binney and Tremaine, 1987, p. 375) called
``swing amplification."\footnote{As was pointed out to us by a second
anonymous referee of the paper, Lau and Bertin (1978) and Lin and
Lau (1979) suggested of $c_r < (2 \Omega / \kappa)
c_{\mathrm{T}}$ as a criterion for
appreciable swing amplification rather than as a criterion for local
nonaxisymmetric instability of gravity perturbations.  The point is
that there exists ambiguity in the interpretation of ``swing" as a
transient temporal amplification of single wavelets, or as a steady
amplification of propagating waves that reflect and form standing
patterns (global normal modes).  Most workers in the field feel more
comfortable with nonaxisymmetric stability criteria that derive from
global normal-mode calculations than local (shearing-sheet) analyses.
Because of collisional damping, the former is inappropriate for
Saturn's rings.}

It is clear from criterion (18) that stability of
{\it nonaxisymmetric} Jeans perturbations, $m$ or $\psi \ne 0$, in 
particular very open perturbations with $\psi \rightarrow 90^\circ$,
in a {\it differentially} rotating disk ($2 \Omega / \kappa > 1$)
requires a larger velocity dispersion than Toomre's critical
value $c_{\mathrm{\small T}}$.  It is only for arbitrary perturbations in
a rigidly rotating disk ($2 \Omega / \kappa = 1$) and/or for
axisymmetric perturbations in a differentially rotating disk 
$c_{\mathrm{crit}} = c_{\mathrm{\small T}}$.  Thus, nonaxisymmetric
Jeans-type disturbances in a nonuniformly rotating system are more 
difficult to suppress than the axisymmetric ones, in general
agreement with the work by Goldreich and Lynden-Bell (1965) and
Julian and Toomre (1966).  The result is quite obvious: spiral
perturbations, in contrast with radial ones, are subject to the
influence of the differential character of the rotational motion.
An expression for
the critical velocity dispersion that is similar to formula (18) 
was first obtained by Lau and Bertin (1978) and  Lin and Lau
(1979, p. 130) in the framework of
the simple hydrodynamical model 
and Morozov (1980, 1981) in more complicated stellar dynamical model.  
The free kinetic energy associated with the differential
rotation of the system under study is only one possible source for
the growth of the energy of these spiral Jeans-type perturbations, and
appears to be released when angular momentum is transferred outward.
According to Polyachenko (1989) and Polyachenko and Polyachenko (1997),
the marginal stability condition
for Jeans perturbations of an arbitrary degree of axial asymmetry
has been available since 1965 (Goldreich and Lynden-Bell, 1965),
though in a slightly masked form.  See Polyachenko and Polyachenko
(1997) for a detailed discussion of the problem.

As one can see from Eq. (18), the modified critical velocity dispersion
$c_r$ grows with $(2 \Omega / \kappa - 1) \propto |d \Omega / d r|$. 
This finding can be regarded as evidence of the fierce spiral Jeans-type
instability of disks with a strong degree of differential rotation: the
shear, $2 \Omega / \kappa > 1$, gives rise to a destabilization effect.

Strictly speaking, the above expression for the critical $c_r$ cannot
be used when the pitch angle is large, since the asymptotic expansion
is valid only for $\tan^2 \psi \ll 1$ (Eq. [7]).  It indicates only a
tendency with increasing $\psi$, and in the case of very open 
spirals with $\psi > 45^\circ$ a special more accurate
analysis, e.g. Polyachenko and Polyachenko (1997), is necessary.  Hence,
the above expression for the critical velocity dispersion is clearly
only approximate.  Note only that according to $N$-body simulations made
by Griv {\it et al.} (1999b) the pitch angle of the most unstable 
Jeans-type perturbations $\psi \approx 35^\circ$, thus $\tan^2 \psi = 
k_\varphi^2 / k_r^2 \ll 1$ and the asymptotic Lin-Shu type approximation
of moderately tightly wound perturbations used throughout the theory
in the present paper does not fail.

The velocity dispersion in a particulate disk is conveniently
expressed in terms of Toomre's (1964) parameter $Q$, which
measures the ratio of actual radial velocity dispersion to the minimum
required to suppress the instability of axisymmetric perturbations:
\be
Q = {c_r \over c_{\mathrm{\small T}}} \equiv 
{c_r \kappa \over 3.36 G \sigma_0} .
\ee
\noindent
It follows from Eqs. (18) and (19) that $Q \approx 2 \Omega / \kappa
\approx 2$ is sufficient to suppress
the instability of arbitrary gravity perturbations
in the disk with Keplerian-like rotation, including the most
unstable open ones.  
Thus, in the differentially rotating disk with rare collisions the
value of $Q$ has to be maintained under about 2 if nonaxisymmetric
Jeans-type instabilities are to be developed.  The latter result has
been obtained in stellar dynamics both analytically (Morozov, 1980,
1981; Griv and Peter, 1996) and by $N$-body simulations (Sellwood and
Carlberg, 1984; Griv, 1998).  Interestingly, Bottema (1993) has found 
that $Q$ between 2 and $2.5$ over a large range of galactic disks. 

Some indirect estimates have probably
indicated the value of $Q \approx 2$ in the B ring of the
Saturnian ring system (Lane {\it et al.}, 1982, p. 543).
So, other Saturn's rings are likely to have the same or even  
smaller $Q$-values, and therefore they may be Jeans-unstable to
spiral Jeans-type perturbations.  According to the above, the
characteristic scale of this instability is $\lambda_{\mathrm{
crit}} = 2 \pi / k_{\mathrm{crit}} \approx 4 \pi \rho$, and
only wavelengths close to $\lambda_{\mathrm{crit}}$ are
unstable.

Notice also that the critical value
of $Q \approx 2$ predicted in our analysis is close to Salo (1992,
1995), Willerding (1992), Osterbart and Willerding (1995),
Sterzik {\it et al.} (1995), and Griv {\it et al.} (1999b) 
numerical results.\footnote{As has been pointed out to us by the referee
of the paper, R. H. Durisen, estimating the velocity dispersion in
different Saturn's ring regions with a steep particle-size 
distribution is not a simple problem: the dispersion velocities are
only ``measured" through analysis of density waves damping.  Also,
there are real differences in the ``opacity" of different ring
regions, e.g. Cuzzi and Estrada (1998).  In this regard, we note
the related idea of DEB's (Dynamic Ephemeral Bodies) in Saturn's
rings (Davis {\it et al.}, 1984; Longaretti, 1989).}

Summarizing, collective motion connected with the Jeans-unstable mode
is excited in the plane of a disk of mutually gravitating particles
when the random velocity dispersion is not sufficiently large, $c_r
< (2 \Omega / \kappa) c_{\mathrm{\small T}}$, 
in other words, if the effective Toomre's
$Q$-value is $Q < 2 \Omega / \kappa \sim 2$.  The characteristic 
local scale of this instability is about $\lambda_{\mathrm{crit}} 
= 4 \pi c_r / \kappa$.  The principal
new contribution of our investigation already presented by Griv
(1996) and Griv and Yuan (1996) is that in the low optical depth
portion of Jeans-stable, differentially rotating, and spatially
inhomogeneous particulate disk with $\tau \simeq \nu_{\mathrm{c}} /
\Omega \ll 1$ there remains however
a Landau-type oscillating microinstability, which is due
to a resonant interaction of particles drifting at the phase velocity
of the nonaxisymmetric Jeans-stable wave with the wave field at
corotation, i.e. \v{C}herenkov radiation effect in plasmas. 
This new kinetic-type, 
$\Re \omega_{* 1,2} \ne 0$, $\Im \omega_{* 1,2} > 0$, and
$|\Im \omega_{* 1,2} / \Re \omega_{* 1,2}| \ll 1$, 
instability of a particulate 
system (or ``a strong resonant version of the diocotron instability")
is fundamentally different from the ordinary aperiodic Jeans
(``diocotron") instability just discussed above.  
The density waves propagating in the plane of a system are effectively
produced at resonant wave-particle
interaction with the growth rate of the mode of maximum instability
$\Im \omega_{* 1,2} \sim 0.1 \Omega$.  The typical radial wavelength of
oscillatory unstable waves $\sim 2 \pi \rho \sim h$.
Rare interparticle collisions cannot suppress this
resonant self-excitation of spiral density waves in Saturn's rings.
The free kinetic energy associated with the differential rotation
is only one possible source for the growth of the average wave energy.
Such kinetic-type instabilities distinguish themselves in that only a
rather small group of so-called resonant particles takes part in 
their generation.  For this reason the energy capacity of the kinetic
instabilities, as a rule, is considerably less than the energy 
capacity of the hydrodynamic Jeans-type instabilities; the latter 
are generated by almost all the particles of the phase space.
To stress, the kinetic Landau-type instability of the
particulate medium may be excited in the gravitating, differentially
rotating, and spatially inhomogeneous disk provided that Jeans
instabilities are already completely suppressed by the combined effect 
of rotation and thermal motions, Eq. (18).

It follows from Eq. (17), the wavelength of maximum Jeans
instability $\lambda_{\mathrm{crit}} = 
2 \pi / k_{\mathrm{crit}}$ is given by
\be
\lambda_{\mathrm{crit}} \approx (2 \Omega / \kappa)
2 \pi \rho \approx 4 \pi c_r / \kappa ,
\ee
\noindent
where $c_r$ is expressed by Eq. (18).  For the parameters of
Saturn's rings $\lambda_{\mathrm{crit}} \sim 100$ m, and
such scales of a few hundred meters correspond to the hyperfine
ringlet structure of Saturn's rings probably already discovered by
Voyager 2 in low optical depth regions.  
This dynamical model may help to explain the observed
quadrupole azimuthal brightness asymmetry of the A ring (see the
Introduction).  

Recently Griv (1998) and Griv {\it et al.} (1999b) studied the
almost collisionless self-gravitating particulate disk 
using the numerical method of local simulations (or $N$-body
simulations in a Hill's approximation).  It was shown that the local
stability criterion obtained from the computer models is in general
agreement with the theoretical prediction as outlined in the present
paper.  Moreover, it has been shown that, as a direct result of the 
Jeans instability of nonaxisymmetric perturbations, the low optical 
depth of such a system is subdivided into numerous irregular
ringlets, with size and spacing of the order $\lambda_{\mathrm{crit}}$.

According to Eq. (17), the Jeans-unstable perturbations
$(\omega_J^2 < 0)$ grow almost aperiodically, so the arbitrary perturbation
therefore does not propagate, but stands still and grows.  To orders
of magnitude the growth rate of this fierce instability 
$$
\Im \omega_{* 1,2} \sim \sqrt{2 \pi G \sigma_0 (k_*^2 / |k|) 
\exp (- k_*^2
\rho^2)} \stackrel{<}{\sim} \Omega ,
$$
\noindent
where $k_*^2 \rho^2 \stackrel{<}{\sim} 1$.
This means that the instability under
investigation will develop rapidly on a dynamical
time scale $\sim 1 / \Omega$.  Inevitably, the velocity
dispersion of particles would be expected to increase in the
field of unstable waves with an amplitude increasing with
time at the expense of the regular circular velocity
(Griv {\it et al.}, 1994).  That is, if at the beginning the
dispersion remains below the critical value $c_{\mathrm{crit}} 
\approx 2 c_{\mathrm{\small T}}$  (or the critical $Q_{\mathrm{crit}}
\approx 2$, respectively) in Saturn's rings, owing to the gravitational
collapses it will increase until this value is reached, i.e. the
situation is stabilized.  The Jeans instability grows on a dynamical
time-scale and heats the disk with rare
interparticle collisions until $Q \approx Q_{\mathrm{crit}}$
(Morozov, 1980, 1981; Griv {\it et al.}, 1999b). 
With the disappearance of the Jeans
instabilities the increase of velocity dispersion will stop. 
The theory of such collective wave-particle interaction was 
initially developed in a theory of plasmas, which is the
quasi-linear theory (Alexandrov {\it et al.}, 1984, p. 408;
Krall and Trivelpiece, 1986, p. 520).  From the plasma theory,
it follows that the characteristic time in which the square of
total spread of random velocities approximately doubles, is
equal to $t_{\mathrm{d}} \sim 1 / \Im \omega_{* 1,2}$.  Therefore,
in the case of the instability $t_{\mathrm{d}}$ is about only the
time of a single revolution of a system $\sim 1/ \Omega$. The
increase of velocity dispersion takes place because the particles gain
additional oscillatory energy in the gravitational field of unstable
density waves (Griv {\it et al.}, 1994).  This nonresonant
process of the heating of the medium by a rise in the
amplitude of the unstable waves growing
almost aperiodically recalls the case of nonresonant
quasi-linear relaxation in plasmas, which can effectively
heat the medium even without interparticle collisions 
(Alexandrov {\it et al.}, 1984; Krall and Trivelpiece,
1986).  Of course, this is an apparent heating and there 
is no change in entropy (Griv {\it et al.}, 1994).
As a result of this increase of the velocity dispersion up to
$c_r \approx 2 c_{\mathrm{\small T}}$ (or $Q \approx 2$, respectively), 
the Jeans instability will be switched off.  Note that such
fast growth in the velocity dispersion has been observed
in $N$-body simulations of Jeans-unstable stellar disks
(Sellwood and Carlberg, 1984; Grivnev 1985; Griv and Chiueh, 1998).
Apparently, Toomre (1964) and Kulsrud
(1972) were the first who discussed an enhancement the rate 
of relaxation of particulate systems toward thermal equilibrium
(or quasi-equilibrium) by collective effects.  That is, relaxation
may occur by collective collisions between one particle and many
others which are collected together by some coherent process such
as a wave (see Kulsrud (1972) for an explanation). 

Thus, in the nonlinear regime, the particles in 
low optical depth regions of Saturn's rings (and the stars
in galaxies) can continue developing Jeans-unstable condensations
only if some effective mechanism of ``cooling" exists.  It seems that
just this approach is assumed in the modern version of the long-lived
spiral structure in numerical models of a stellar-gaseous galactic
disk --- suggesting the dissipation in the gas and accretion, and
formation of new dynamically cold stars (Sellwood and Carlberg, 1984;
Griv and Chiueh, 1998).  
Obviously, in Saturn's rings such cooling mechanism is
actually operating: inelastic/dissipative interparticle collisions
reduce the magnitude of the relative velocity of particles.
By local $N$-body simulations, Salo (1992, 1995), Richardson (1994), 
and Osterbart and 
Willerding (1995) already 
investigated the role of the Jeans instability type mechanism in
long-term sculpting of the structure of
Saturn's rings by including both gravitational interactions and
dissipative impacts between particles.  It seems likely that this
cooling mechanism might operate effectively in the low-optical
depth regions of the Saturnian ring
system, and might play an important role in the development of a
hyperfine $\sim 4 \pi \rho$ long-lived ring structure therein.

In turn, in the another frequency range
$$
|\omega_*| \sim \left| \omega_{\mathrm{gr}} \right| \ll
|\omega_{\mathrm{J}}| 
\quad \mbox{and} \quad |\omega_*| \gg \nu_{\mathrm{c}} ,
$$
\noindent
the dispersion relation (14) has also another root equal to
$$
\omega_{* 3} \approx \omega_{\mathrm{gr}} {\kappa^2 \over 
\omega_{\mathrm{J}}^2}
- i \nu_{\mathrm{c}} .
$$
\noindent
This root describes the gradient, $L^{-1} \ne 0$, branch of oscillations
modified by collisions.
Apparently, the gradient perturbations are stable and are
independent of the stability of Jeans modes. 
These low-frequency, $|\omega_{* 3}| \ll \Omega$, gradient
waves are weakly damped $(\Im \omega_{* 3} < 0)$ so that in the plane
of a collisional disk weakly damped small-amplitude waves can propagate.

\subsection{The dissipative oscillations --- strong and frequent collisions 
($\omega^2 \ll \nu_{\mathrm{c}}^2$ and $\nu_{\mathrm{c}}^2 \gg \Omega^2$)}

Such high optical depth regions, in which one cannot treat collisions 
as a small perturbation, can be found in the central portions of the
B ring at distances 
$100 \, 000 < r < 122 \, 000$ km from Saturn's center, where
$\tau > 1$ (Esposito, 1986, Fig. 2).  This is just the procedure in
ordinary gas dynamics, where collisions are the dominant effect
and the mean potential between neighboring particles is large
compared with the thermal energy.
This is the usual hydrodynamic approximation accepted by Lynden-Bell
and Pringle (1974), Morozov {\it et al.} (1985), Willerding (1992),
and Schmit and Tscharnuter (1995, 1999) for
perturbations with small epicyclic radius, $k_* \rho \le 1$ (and
$\nu_{\mathrm{c}}^2 \gg \kappa^2$).  

In the limit when the frequency of oscillations is smaller than the
collision frequency and the collisional frequency is greater than the
rotational (epicyclic) frequency, the effect of the system's rotation is
negligible.  In a plasma, this limit corresponds to the isotropic plasma
without an external magnetic field.  In such a nonrotating system, in
the lowest approximation of the theory the particle dynamics and wave
propagation properties in a system correspond to an unmagnetized plasma
and free-streaming particle orbits
$$
\vec{r}^\prime = \vec{r} + \vec{v} (t^\prime - t) 
\quad \mbox{and} \quad
\vec{v}^\prime = \vec{v} ,
$$
\noindent
where $\vec{r}$ is the radius-vector of a particle and $\vec{v}$
is the particle's velocity.  This is physically obvious since a typical
particle should follow at least one epicycle between collisions.  Only
in this limit can one speak of the epicycle rotation.  The equilibrium
axially symmetric distribution function is the Maxwellian with the 
``temperature" $c^2$ and the surface density $\sigma_0$:
$$
f_0 ( |\vec{v}| ) = {\sigma_0 \over 2 \pi c^2} \exp \left( -
{v^2 \over 2 c^2} \right) .
$$

The formal transition to the limit $\kappa \rightarrow 0$ in Eq. (13)
is nontrivial and connected with the problem of the asymptotic 
representation of the Bessel functions of high order at large arguments.
For $\kappa \rightarrow 0$, the arguments of the Bessel functions in Eq.
(13) become large.  Then, all terms with $|l| < l_{\mathrm{max}} \approx
k_* v_\perp / \kappa$ contribute to the same order, whereas the terms
with $|l| > l_{\mathrm{max}}$ are exponentially small.  Therefore, the
summation in Eq. (13) must be extended up to $|l| = l_{\mathrm{max}}$.
Such a solution has been attempted in plasma physics, however, the
calculation was tedious.  A more well-defined approach is to integrate
the basic Boltzmann equation (1) over the free-streaming orbits given 
above when the collision term becomes the principal one,
and obtain a dispersion relation in this ``rotationless" case.
Then, the resulting Boltzmann equation for the perturbed distribution
function of particles (in a rotating frame we are using), 
$$
{\partial f_1 \over \partial t} + \vec{v} \cdot {\partial \Phi_1 \over
\partial \vec{r}} - {\partial \Phi_1 \over \partial \vec{r}} \cdot
{\partial f_0 \over \partial \vec{v}} = - \nu_{\mathrm{c}} f_1 ,
$$
\noindent
can be solved by successive approximations, neglecting the influence
of the first two terms on the left-hand side $\propto \partial /
\partial t$ and $\partial / \partial \vec{r}$ on the distribution of
particles in the zero-order approximation (cf. Rukhadze and Silin
(1969) and Alexandrov {\it et al.} (1984, pp. 66 and 101)).\footnote{The
above is analogous to the efficient Chapman-Enskog method of
solution of the Boltzmann equation for a monatomic gas (Lifshitz
and Pitaevskii, 1981).}  As a result in the zero-order
approximation, one gets
\be
f_1 = i {\Phi_1 \vec{k} \over \nu_{\mathrm{c}}} \cdot {\partial f_0 \over
\partial \vec{v}} .
\ee

The perturbed distribution $f_1$ given by Eq. (21) can be used in the
continuity equation to determine perturbation of the surface density.
The continuity equation for a small density perturbation $\sigma_1
(\vec{r}, t)$ in a disk is
\be
\sigma_1 = - \int_{-\infty}^t d t^\prime {\partial \over \partial
\vec{r}} \int_{-\infty}^\infty
\vec{v} f_1 d \vec{v} ,
\ee
\noindent
where $|\sigma_1 / \sigma_0| \ll 1$ and $\sigma_1 = 0$ 
as $t \rightarrow - \infty$.  The fact that the density of the
disk can vary only as a result of diffusion of particles in the disk,
i.e. any external  (or nondiffusion) fluxes of matter are absent, has
taken into account in the equation of continuity.  Utilizing in Eq. (22)
the above equation for the $f_1$, we thus arrive at the following formula 
for the perturbed surface density in a spatially homogeneous disk:
\be
\sigma_1 = {\Phi_1 k^2 \over i \nu_{\mathrm{c}}} \int_{-\infty}^\infty \vec{v}
\cdot {\partial f_0 \over \partial \vec{v}}{d \vec{v} \over \vec{k}
\cdot \vec{v} - \omega_* - i \epsilon} ,
\ee
\noindent
where the positive infinitesimal $\epsilon$ was introduced, 
which serves to assure the adiabatic turning on the perturbation
(Lifshitz and Pitaevskii, 1981; Alexandrov {\it et al.}, 1984).
Following Lifshitz and Pitaevskii (1981, p. 121), we take the
$r$-axis along $\vec{k}$.  Then
\be
\sigma_1 = {\Phi_1 k_r^2 \over i \nu_{\mathrm{c}}} \int_{-\infty}^\infty
v_r {\partial f_0 (v_r) \over \partial v_r}{d v_r \over 
k_r v_r - \omega - i \epsilon} ,
\ee
\noindent
where we used the distribution function only with respect to
$v_r$:
$$
f_0 (v_r) = \int_{-\infty}^\infty f_0 d v_\varphi .
$$

In Eq. (24), by considering the most interesting high-frequencies
perturbations when $\omega^2 \gg k_r^2 v_r^2$, the integral may be 
evaluated by using the binominal theorem:
$$
\int_{-\infty}^\infty v_r {\partial f_0 (v_r) \over \partial v_r}
{d v_r \over \omega - k_r v_r} = \int_{-\infty}^\infty v_r {\partial
f_0 (v_r) \over \partial v_r}{d v_r \over \omega} \left( 1 + {k_r v_r
\over \omega} + {k_r^2 v_r^2 \over \omega^2} + \cdots \right) .
$$
\noindent
The integrals of the terms even in $\omega$ are zero; the remainder
give
\be
\sigma_1 \approx {\Phi_1 k_r^2 \sigma_0 \over i \nu_{\mathrm{c}} \omega} 
\left(
1 + {3 k_r^2 c_r^2 \over \omega^2} \right) ,
\ee
\noindent
where $k_r^2 c_r^2 / \omega^2 \ll 1$.
 
A comparison of Eqs. (7) and (25) leads to the sought dispersion
relation in the fast wave range where the phase velocity exceeds the
thermal velocities $\omega / k_r > c_r$,
\be
1 + {2 \pi G \sigma_0 |k_r| \over i \nu_{\mathrm{c}} \omega} 
\left( 1 + {3 k_r^2
c_r^2 \over \omega^2} \right) = 0 ,
\ee
\noindent 
which finally yields
\be
\omega = i {2 \pi G \sigma_0 |k_r| \over \nu_{\mathrm{c}}} 
\left[ 1 - 3 \nu_{\mathrm{c}}^2
\left( {c_r \over 2 \pi G \sigma_0} \right)^2 \right] .
\ee
\noindent
We have took into account that $\omega \approx i 2 \pi G \sigma_0
|k_r| / \nu_{\mathrm{c}}$ and substituted this value for $\omega$ 
in a small
term in Eq. (26).  Due to condition $\omega^2 / k_r^2 \gg c_r^2$, the
second term on the right-hand side in Eq. (27) is a small correction.

Equation (27) gives almost the same condition
for secular dissipative-type instability as in the usual hydrodynamic
description adopted by Lynden-Bell and Pringle (1974), Fridman and
Polyachenko (1984, Vol. 2), 
Morozov {\it et al.} (1985), and others: a disk with frequent
collisions is always aperiodically unstable for all wavelengths.
Frequent binary collisions thus remove the rotational stabilization in
a flat system.  This fundamental result was found first by Lynden-Bell 
and Pringle (1974).  Contrary to Lynden-Bell and Pringle, Fridman and
Polyachenko, Morozov {\it et al.}, and others, the result (27) obtained
in the present paper through the use of the kinetic approximation states
that the given instability in a particulate system develops for all modes,
but not only for modes with wavelengths longer than a critical one.
To orders of magnitude the growth rate of the instability in Saturn's
rings $\Im \omega = (0.3-0.5) \Omega$.  This is because in the B ring
$2 \pi G \sigma_0 |k_r| \sim \Omega^2$ and 
$\nu_{\mathrm{c}} / \Omega = 2 - 3$.
Thus, the instability will develop rapidly on a time scale of $3-5$
rotations.
Interestingly, the growth rate of secular-unstable perturbations
decreases with increasing 
$\nu_{\mathrm{c}}$ and $c_r^2$.  This is physically
obvious: as a result of interparticle collisions and thermal pressure, 
the organized motion tends to be lost.  Similar to the case of
the classical Jeans instability considered in the previous subsection,
it may be shown that the inclusion of spatial inhomogeneity leads to
a small real part of the wavefrequency $\omega$.  Like the Jeans 
instability, the secular dissipative-type instability presumably heats
the disk with frequent interparticle collisions.  According to the
equation above, the latter leads to a decrease in the growth rate of
the oscillation amplitude.  Eventually, as a result of such ``heating"
the secular instability will be switched off.  It seems likely that,
in the nonlinear regime, the particles in Saturn's rings can continue
developing dissipative-unstable condensations: inelastic physical
collisions reduce the magnitude of the relative
velocity of particles, and, thus, effectively cool the system.

There are similarities between the aforementioned almost aperiodic
instability in a particulate disk with frequent collisions and
the secular dissipative-type
instability in a rigidly rotating gaseous viscous sheet that was first
discovered by Lynden-Bell and Pringle (1974).  They claimed this
instability to be analogous to the well-known viscous mechanism that
converts Maclaurin spheroids to Jacobi ellipsoids.  (In fact,
Lynden-Bell and Pringle derived the dispersion relation for a rigidly 
rotating sheet.)  Fridman and Polyachenko (1984, Vol. 2, p. 239),
Morozov {\it et al.} (1985), and Willerding (1992) improved the 
Lynden-Bell and
Pringle calculation by taking into account the effect of nonuniform
rotation in a two-dimensional self-gravitating system.  
Fridman and Polyachenko showed that the perturbed, sliced state
is energetically preferable.  Fridman and Polyachenko, and Morozov
{\it et al.}
explained the cause of the instability: it results from perturbations
which have negative energy in the dissipative medium.  Thus, this
instability might arise merely from the dissipation of the energy of
regular circular rotation into ever larger amounts of heat, i.e.
the energy of medium's regular motion transforms into random motions 
of particles.  According to 
Morozov {\it et al.} (1985) and Eq. (27) the introduction of 
differential rotation leaves the result of Lynden-Bell and Pringle
unchanged.  Similar studies of the secular dissipative-type ring
instability in the framework of the simple hydrodynamic model were
made also by Schmit and Tscharnuter (1995, 1999). 
To emphasize, the instability
under consideration, which has an essential dependence on the
self-gravitation of the disk matter, will remain even in a rigidly
rotating disk.  Gorkavyi {\it et al.} (1990) attempt to apply the
secular dissipative-type instability to the distance law of planets.
By a particle $N$-body simulations, Sterzik {\it et al.} (1995)  
probably confirmed predictions regarding wavelength and growth time  
of the instability.  Sterzik {\it et al.} especially stressed that
any dissipative mechanism (e.g. convectively or magnetically or
shear-generated turbulence, inelastic particle collisions) can cause
the instability, as long as it reflects a hydrodynamic shear viscosity.

Equation (27) only indicates that the growth rate is a maximum for
short radial wavelength, $\lambda_r = 2 \pi / k_r \rightarrow 0$, but
cannot determine the region of maximally unstable wavelengths.  This
actually means that the characteristic scale of stratification of the
disk to which the secular instability leads is unknown.  
However, we have
restricted ourselves to the consideration of relatively long-wavelengths
modes with $\lambda_r \gg h/2$.\footnote{Small-scale perturbations
with $\lambda_r < h$ are stable (Fridman and Polyachenko, 1984).}
We conclude that, as in the case of rare and weak collisions, the
instability growth rate is maximum for the wavelengths $\lambda_r 
\sim (2-4) \pi h$, that is, the growth rate is a maximum for 
perturbations of the order the Jeans-Toomre wavelength, $\sim
\lambda_J = c^2 / G \sigma_0$.  Here, in accordance with the 
results of the previous subsection, we took into account that in
marginally Jeans-stable disks $c \approx (2 \Omega / \kappa) 
c_{\mathrm{\small T}}$.
In the hydrodynamical model of a galactic gaseous disk, Morozov 
{\it et al.} (1985), and in the hydrodynamical model of Saturn's ring 
disk, Schmit and Tscharnuter (1995, 1999) already
pointed to the most unstable perturbations $\approx 4 \lambda_J$ in
marginally Jeans-stable parts of a self-gravitating system with 
frequent interparticle collisions. 

To emphasize, the instability has an essential dependence on the
self-gravitation of the disk matter, and it has nothing to do with
a ``purely viscous" mechanism suggested by
Lin and Bodenheimer (1981), Lukkari (1981), and Ward (1981)
to describe the irregular ringlets in regions of high
optical depth around Saturn by taking into account the effects of
viscous forces, but neglecting self-gravity.\footnote{Araki and
Tremaine (1986), Wisdom and Tremaine (1986), Brophy and Esposito
(1989), and Richardson (1994) have proved that the ``purely viscous" 
type of instability would not actually occur in Saturn's rings.}

\section{Conclusions}

Collective-type oscillations (and their stability) of highly flattened
self-gravitating systems have been investigated both experimentally by
gravitational $N$-body simulations and theoretically for more than
three decades.  The linear theory of such oscillations is well known.
In this paper the linear kinetic theory of oscillations and their   
stability of the highly flattened, rapidly (and nonuniformly) rotating 
disk of mutually gravitating particles is reexamined and extended. 
The simultaneous effect of self-gravity and collisions between
particles is taken into account.  The stability analysis 
of small gravity perturbations is
based on the Boltzmann kinetic equation with the Krook
model collision integral, modified in accord with Shu and Stewart
(1985) to allow for the effects of inelastic collisions between
particles.  Interest in the kinetic theory is due to the efforts
to solve the problem of small-scale irregular radial structure in  
Saturn's rings probably already revealed by the Voyager mission.

It is found that in the two limiting cases of weak and rare and 
strong and frequent physical inelastic collisions
the particulate disk may be unstable 
with respect to the Jeans-type perturbations and to the secular
dissipative-type perturbations, respectively.  These instabilities
of small-amplitude gravity perturbations (e.g. those produced by a
spontaneous perturbation and/or a companion system) do not depend on
the behavior of the particle distribution function in the neighborhood
of a particular speed, but the determining factors of the instabilities
are the macroscopic parameters of a self-gravitating system 
--- the mean mass density,
the angular velocity of regular rotation, the dispersion of random
velocities of particles, and the frequency of interparticle collisions.  
Generally, the growth rate of these almost aperiodic ($|\Re \omega /
\Im \omega| \ll 1$) nonresonant instabilities is large, $\Im \omega 
\sim \Omega$, where $\omega$ is the frequency of excited waves.
Thus, such oscillatory unstable
perturbations will develop rapidly on the time scale of only
a few revolutions of the system under study.
The typical wavelength of the most unstable perturbations is
$\lambda_{\mathrm{crit}} \approx 4 \pi \rho \approx 2 \pi h$.
For the parameters of Saturn's rings $\lambda_{\mathrm{crit}} 
\sim 100$ m.

We suggest that the instabilities we are investigating may be
considered as an effective generating mechanism for hyperfine 
structure of the order of one Jeans length, i.e. $\sim 4 \pi \rho 
\approx 2 \pi h$ in the main C, A, and B rings of the Saturnian ring 
system.  Because modern observations indicate that the ring
thickness ranges from 1 m or less in the C ring to $1-5$ m in the B
ring and $5-30$ m in the A ring (Esposito, 1986, 1993), more accurate
estimations give the critical value of wavelength
$\lambda_{\mathrm{crit}}$ in the C ring $\lambda_{\mathrm{crit}}
\stackrel{<}{\sim} 7$ m, in the B ring $\lambda_{\mathrm{crit}} 
= 7-30$ m, and in the A ring $\lambda_{\mathrm{crit}} = 30-200$ m.

In addition, we can expect the formation of small-scale spiral arm
and/or radial fragments as precursors of small ``moonlets" inside the
Roche limit in the Saturnian ring system and protoplanetary disks of
particles.  The latter is an important step towards an understanding
of a main question of protoplanetary disk evolution as well as the 
evolutionary processes in galactic disks: what kind of evolutionary
processes leads to the formation of moons, planets, and stars in
different astrophysical disk system?

As has been pointed out in the Introduction, by Schmit and Tscharnuter 
(1995) and Griv (1996, 1998), structures on the $100$ m scale fall
below the noise of Voyager's stellar occultation data.  More precise
Cassini spacecraft observations should help to settle this question
almost 6 years from now, but at present the paper would have to be
regarded as a prediction of structure in ring systems based on 
theoretical modeling rather than as an explanation for observed features.

Further work is in progress to extend the calculations presented in the
article.  In addition, it would be desirable to have experimental tests
($N$-body experiments) of the theory.

\bigskip\noindent
{\it Acknowledgements.}
The authors wish to thank numerous colleagues for many discussions,
especially Kai-Wing Chen, Tzihong Chiueh, Alexei M. Fridman, Peter
Goldreich, Edward Liverts, Muzafar N. Maksumov, William Peter, Shlomi
Pistinner, Frank H. Shu, and Raphael Steinitz.  We are grateful
to the first referee of the paper, R. H. Durisen, for very useful
comments and constructive criticism that improved our understanding
of the problems under consideration.  This research was sponsored
in part by the Ministry of Science and the Science Foundation 
founded by the Academy of Sciences and Humanities in 
Israel, and the Academia Sinica in Taiwan.  One of us (E. G.)
was finantially supported by the Israeli Ministry of Immigrant
Absorption in the framework of the program ``Giladi."

\begin{center}
{\bf References}
\end{center}

\noindent
Alexandrov, A. F., Bogdankevich, L. S., and Rukhadze, A. A. (1984)
  {\it Principles of Plasma Electrodynamics}. Springer, Berlin.\\
Araki, S. (1991) Dynamics of planetary rings.
  {\it Am. Scientist} {\bf 79}, 44-59.\\
Araki, S. and Tremaine, S. (1986) 
  The dynamics of dense particle disk. 
  {\it Icarus} {\bf 65}, 83-109.\\
Bertin, J. (1980) 
  On the density wave theory for normal spiral galaxies.
  {\it Phys. Rep.} {\bf 61}, 1-69.\\
Bertin, J. and Mark, J. W.-K. (1978) 
  Density wave theory for spiral galaxies: the regime of finite
  spiral arm inclination in stellar dynamics.
  {\it Astron. Astrophys.} {\bf 64}, 389-397.\\
Bhatnagar, P. L., Gross, E. P., and Krook, M. (1954) 
  Model for collisional processes in gases.
  {\it Phys. Rev.} {\bf 94}, 511-527.\\
Binney, J. and Tremaine, S. (1987) {\it Galactic Dynamics}.
  Princeton Univ. Press, Princeton.\\
Borderies, N. (1989) Ring dynamics.
  {\it Celes. Mech. Dynamical Astr.} {\bf 46}, 207-230.\\
Bottema, R. (1993) 
  The stellar kinematics of galactic disks.
  {\it Astron. Astrophys.} {\bf 275}, 16-36.\\ 
Bridges, F. G., Hatzes, A. P., and Lin, D. N. C. (1984)
  Structure, stability, and evolution of Saturn's rings.
  {\it Nature} {\bf 309}, 333-335.\\
Brophy, T. G. and Esposito, L. W. (1989) 
  Simulation of collisional transport processes and the stability
  of planetary rings.  {\it Icarus} {\bf 78}, 181-205.\\
Brophy, T. G. and Rosen, P. A. (1992)
  Density waves in Saturn's rings probed by radio and and optical 
  occultation: Observational tests of theory.
  {\it Icarus} {\bf 99}, 448-467.\\ 
Colombo, G., Goldreich, P., and Harris, A. (1976) 
  Spiral structure as an explanation for the asymmetric brightness
  of Saturn's A Ring.  {\it Nature} {\bf 264}, 344-347.\\
Colwell, J. E. (1994) 
  The disruption of planetary satellites and the creation of 
  planetary rings.
  {\it Planet. Space Sci.} {\bf 42}, 1139-1149.\\ 
Contopoulos, G. (1979) 
  Inner Lindblad resonance in galaxies.  Nonlinear theory.
  {\it Astron. Astrophys.} {\bf 71}, 221-244.\\
Cuzzi, J. N. (1989) Saturn: Jewel of the solar system.
  {\it Planetary Rep.} {\bf 9}, 12-15.\\
Cuzzi, J. N. and Burns, J. A. (1988) 
  Charged particle depletion surrounding Saturn's F ring:
  evidence for a moonlet belt?
  {\it Icarus} {\bf 74}, 284-324.\\
Cuzzi, J. N., Lissauer, J. J., and Shu, F. H. (1981)
  Density waves in Saturn's rings.
  {\it Nature} {\bf 292}, 703-707.\\
Cuzzi, J. N., Lissauer, J. J., Espositi, L. W., Holberg, J. B., Marouf,
  E. A., Tyler, G. L., and Boischot, A. (1984) 
  Saturn's Rings: Properties and Processes. In {\it Planetary Rings}
  (edited by Greenberg, R. and Brahic, A.), pp. 73-199.  Univ. of
  Arizona, Tucson.\\ 
Cuzzi, J. N. and Estrada, P. R. (1998) 
  Compositional Evolution of Saturn's Rings Due to Meteoroid Bombardment.
  {\it Icarus} {\bf 132}, 1-35.\\
Davidson, R. C. (1992) {\it Physics of Nonneutral Plasmas}.
   Addison-Wesley, Redwood City, CA.\\
Davis, D. R., Weidenschilling, S. J., Chapman, C. R., and Greenberg, R.
  (1984) Saturn ring particles as dynamic ephemeral bodies.
  {\it Science} {\bf 224}, 744-747.\\ 
Dones, L. (1991) 
  A recent cometary origin for Saturn's rings? 
  {\it Icarus} {\bf 92}, 194-203.\\
Dones, L., Cuzzi, J. N., and Showalter, M. R. (1993)
  Voyager photometry of Saturn's A ring.
  {\it Icarus} {\bf 105}, 184-215.\\
Durisen, R. H., Cramer, N. L., B. W. Murphy {\it et al.} (1989)
  Ballistic transport in planetary ring systems due to particle
  erosion mechanisms.  I. Theory, numerical methods and
  illustrative examples.  {\it Icarus} {\bf 80}, 136-166.\\
Durisen, R. H., Bode, P. W., S. J. Dyck {\it et al.} (1996)
  Ballistic transport in planetary ring systems due to particle 
  erosion mechanisms.  {\it Icarus} {\bf 124}, 220-236.\\
Esposito, L. W. (1986) 
  Structure and evolution of Saturn's rings.
  {\it Icarus} {\bf 67}, 345-357.\\
Esposito, L. W. (1992)
  Running rings around modellers. 
  {\it Nature}, {\bf 360}, 531-532.
Esposito, L. W. (1993) 
  Understanding planetary rings.
  {\it Annu. Rev. Earth Planet. Sci.}
  {\bf 21}, 487-523.\\
Esposito, L. W., Harris, C. C., and Simmons, K. E.
  Features in Saturn's rings. {\it Astrophys. J. Suppl.}
  {\bf 63}, 749-770.\\
Flynn, B. C. and Cuzzi, J. N. (1989) Regular structure
  in the inner Cassini Division of Saturn's rings.
  {\it Icarus} {\bf 82}, 180-199.\\
Franklin, F. A., Colombo, G., and Cook, A. F. (1982)
  A possible link between the rotation of Saturn and its ring 
  structure.  {\it Nature} {\bf 295}, 128-130.\\
Franklin, F. A. and Colombo, G. (1978) 
  On the azimuthal brightness variations of Saturn's rings.
  {\it Icarus} {\bf 33}, 279-287.\\
Franklin, F. A., Cook II, A. F., R. T. F. Barrey {\it et al.} (1987)
  Voyager observations of the azimuthal brightness variations
  in Saturn's rings.  {\it Icarus} {\bf 69}, 280-296.\\
Fridman, A. M. and Polyachenko, V. L. (1984) {\it Physics of 
  Gravitating Systems}, Vols 1 and 2.  Springer, New York.\\
Ginzburg, I. F., Polyachenko, V. L., and Fridman, A. M. (1972)
  Ring stability of Saturn.
  {\it Soviet Astr.} {\bf 15}, 643-647.\\
Goldreich, P. and Lynden-Bell, D. (1965) 
  Spiral arms as sheared gravitational instabilities.
  {\it Mon. Not. R. Astron. Soc.} {\bf 130}, 125-158.\\
Goldreich, P. and Tremaine, S. (1978) 
  The formation of the Cassini Division in Saturn's rings.
  {\it Icarus} {\bf 34}, 240-253.\\
Goldreich, P. and Tremaine, S. (1982)
  The dynamics of planetary rings.
  {\it Annu. Rev. Astr. Astrophys.} {\bf 20}, 249-283.\\
Goldreich, P., Rappaport, N., and Sicardy, B. (1995)
  Single-sided shepherding.
  {\it Icarus} {\bf 118}, 414-417.\\
Gorkavyi, N. N., Polyachenko, V. L., and Fridman, A. M. (1990)
  Dissipative instability of the protoplanetary disk and law
  of planetary distances.
  {\it Soviet Astr. Lett.} {\bf 16}, 79-82.\\
Gresh, D. L., Rosen, P. A., Tyler, G. L., and Lissauer, J. J.
  An analysis of bending waves in Saturn's rings using Voyager
  radio occultation data.  {\it Icarus} {\bf 68}, 481-502.\\
Griv, E. (1996) 
  Resonant excitation of density waves in Saturn's rings.
  {\it Planet. Space Sci.} {\bf 44}, 579-587.\\
Griv, E. (1998) 
  Local stability criterion for the Saturnian ring system.
  {\it Planet. Space Sci.} {\bf 46}, 615-628.\\
Griv, E. and Peter, W. (1996) 
  Stability of the stellar disks of flst galaxies.  
  I. A collisionless, homogeneous system.
  {\it Astrophys. J.} {\bf 469}, 84-98.\\
Griv, E. and Yuan, C. (1996) 
  Resonant excitation of density waves in Saturn's rings:
  the effect of interparticle collisions.
  {\it Planet. Space Sci.} {\bf 44}, 1185-1190.\\
Griv, E. and Yuan, C. (1997) Disk-halo interaction in flat 
  galaxies of stars.
  In {\it Dark and Visible Matter in Galaxies} (edited by Persic, M.
  and Salucci, P.), pp. 228-235.  ASP Conference Series, New York.\\
Griv, E., Chiueh, T., and Peter, W. (1994) 
  The weakly nonlinear theory of density waves in a stellar disk.
  {\it Physica A} {\bf 205}, 299-306.\\
Griv, E. and Chiueh, T. (1997) 
  Secular instability of Saturn's rings.
  {\it Astron. Astrophys.} {\bf 311}, 1033-1042.\\
Griv, E., Yuan, C., and Chiueh, T. (1997a) 
  Resonant excitation of density waves in
  Saturn's rings: the effect of finite disk thickness.
  {\it Planet. Space Sci.} {\bf 45}, 627-635.\\
Griv, E., Gedalin, M., and Yuan, C. (1997b) 
  Collisional dynamics of the Milky Way.  
  {\it Astron. Astrophys.} {\bf 328}, 531-543.\\
Griv, E. and Chiueh, T. (1998) 
  Central NGC 2146: a firehose-type bending instability in 
  the disk of newly formed stars?
  {\it Astrophys. J.} {\bf 503}, 186-211.\\
Griv, E., Yuan, C., and Gedalin, M. (1999a) 
  Small-amplitude density waves in galactic discs with  
  radial gradients.
  {\it Mon. Not. R. Astron. Soc.} {\bf 307}, 1-23.\\
Griv, E., Rosenstein, B., Gedalin, M., and Eichler, D. (1999b) 
  Local stability criterion for a gravitating disk of stars.
  {\it Astron. Astrophys.} {\bf  347}, 821-840.\\ 
Grivnev, E. M. (1985) 
  Development of spiral density waves
  and evolution of their parameters: a computer experiment.
  {\it Soviet Astr.} {\bf 29}, 400-403.\\
Grivnev, E. M. (1988 )
  Galactic star orbits in the post-epicyclic approximation.
  {\it Soviet Astr.} {\bf 32}, 139-143.\\
Hanninen, J. and Salo, H. (1995) 
  Formation of isolated narrow ringlets by a single satellite.
  {\it Icarus} {\bf 117}, 435-438.\\
Hohl, F. (1978) 
  Three-dimensional galaxy simulations.
  {\it Astron. J.} {\bf 83}, 768-778.\\
Holberg, J. B., Forrester, W. T., and Lissauer, J. J. (1982)
  Identification of resonance features within the rings of Saturn.
  {\it Nature}, {\bf 297}, 115-120.\\
Horne, L. J. and Cuzzi, J. N. (1996) 
  Characteristic wavelengths of irregular structure in Saturn's B Ring. 
  {\it Icarus} {\bf 119}, 285-310.\\
Horne, L. J., Showalter, M. R., and Russell, C. T. (1996)
  Detection and behavior of Pan wakes in Saturn's A ring.
  {\it Icarus} {\bf 124}, 663--676.\\
Julian, W. H. and Toomre, A. (1966) 
  Non-axisymmetric response of differentially rotating disks of stars.
  {\it Astrophys. J.} {\bf 146}, 810-830.\\
Kolvoord, R. A. and Burns, J. A. (1992)
  Three-dimensional perturbations of particles in a narrow 
  {\it Icarus} {\bf 95}, 253-264.\\
Krall, N. A. and Trivelpiece, A. W. (1986) {\it Principles of Plasma
  Physics}. San Francisco Press, San Francisco.\\
Kulsrud, R. M. (1972) Enhancement of relaxation processes by
  collective effects. In {\it Gravitational $N$-body Problem}
  (edited by Lecar, M.), pp. 337-346.  Reidel, Dordrecht.\\
Lane, A. L., Hord, C. W., R. A. West {\it et al.} (1982) 
  Photopolarimetry from Voyger 2: preliminary results on Saturn,
  Titan, and the rings.  {\it Science} {\bf 215}, 537-543.\\
Lau, Y. Y. and Bertin, G. (1978) 
  Discrete spiral modes, spiral waves,
  and the local dispersion relation.
  {\it Astrophys. J.} {\bf 226}, 508-520.\\
Lifshitz, E. M. and Pitaevskii, L. P. (1981)
  {\it Physical Kinetics}. Pergamon, Oxford.\\
Lin, D. C. N. and Bodenheimer, P. (1981) 
  On the stability of Saturn's rings.
  {\it Astrophys. J.} {\bf 248}, L83-L86.\\
Lin, C.C. and Shu, F. H. (1966) 
  On the spiral structure of disk
  galaxies. II. Outline of a theory of density waves.
  {\it Proc. Natl. Acad. Sci.} {\bf 55}, 229-232.\\
Lin, C. C. and Lau, Y. Y. (1979) 
  Density wave theory of spiral structure of galaxies.
  {\it Stud. Appl. Math.} {\bf 60}, 97-163.\\
Lin, C. C., Yuan, C., and Shu, F. H. (1969) 
  On the spiral structure of
  disk galaxies. III. Comparison with observations.
  {\it Astrophys. J.} {\bf 155}, 721-745.\\
Lissauer, J. J. (1985) 
  Bending waves and the structure of Saturn's rings.
  {\it Icarus} {\bf 62}, 433-447.\\
Lissauer, J. J. (1989) 
  Spiral waves in Saturn's rings.
  In {\it Dynamics of Astrophysical Discs} (edited by
  Sellwood, J. A.), pp. 1-16. Cambridge Univ. Press, Cambridge.\\
Lissauer, J. J., Shu, F. H., and Cuzzi, J. N. (1981) 
  Moonlets in Saturn's rings? 
  {\it Nature} {\bf 292}, 707-711.\\
Lissauer, J. J. and Cuzzi, J. N. (1985) 
  Rings and moons: clues to understanding of solar nebula. 
  In {\it Protostars and Planets. II}
  (edited by Black, D. C. and Matthews, M. S.), pp. 920-927. 
  Univ. Arizona Press, Tucson.\\
Longaretti, P.-Y. (1989) 
  Saturn's main ring particle size distribution: 
  An analytic approach. {\it Icarus} {\bf 81}, 51-73.\\
Lukkari, J. (1981) 
  Collisional amplification of density fluctuations in 
  Saturn's rings.
  {\it Nature} {\bf 292}, 433-435.\\
Lynden-Bell, D. and Pringle, J. E. (1974) 
  The evolution of viscous discs and the origin of the 
  nebular variables.
  {\it Mon. Not. R. Astron. Soc.} {\bf 168}, 603-637.\\
Lynden-Bell, D. and Kalnajs, A. J. (1972) 
  On the generating mechanism of spiral structure.
  {\it Mon. Not. R. Astron. Soc.} {\bf 157}, 1-30.\\
Mikhailovskii, A. B. (1974) {\it Theory of Plasma Instabilities},
  Vols 1 and 2. Consultants Bureau, New York.\\
Mikhailovskii, A. B. and Pogutse, O. P. (1966) 
  Kinetic theory of the oscillations
  of an inhomogeneous plasma with collisions.
  {\it Soviet Phys.-Tech. Phys.} {\bf 11}, 153-161.\\
Morozov, A. G. (1980) 
  On the stability of an inhomogeneous disk of stars.
  {\it Soviet Astr.} {\bf 24}, 391-394.\\
Morozov, A. G. (1981) 
  On the mass ratio of the galactic halo and disk. 
  {\it Soviet Astr.} {\bf 25}, 421-426.\\
Morozov, A. G., Torgashin, Yu. M., and Fridman A. M. (1985)
  Turbulent viscosity in a gravitating gaseous disk.
  {\it Soviet Astr. Lett.} {\bf 11}, 94-97.\\
Nicholson, P. D. and Dones, L. (1991) 
  Planetary rings.
  {\it Rev. Geophys. Suppl.} {\bf 29}, 313-327.\\
Osterbart, R. and Willerding, E. (1995) 
  Collective processes in planetary rings.
  {\it Planet. Space Sci.} {\bf 43}, 289-298.\\
Pitaevskii, L.P. (1963)
  Effect of collisions on the disturbances around
  a body moving in a plasma.  
  {\it Soviet Phys.-JETP} {\bf 17}, 658-664.\\
Polyachenko, V. L. (1989) 
  Stability criteria for gravitating discs.
  In {\it Dynamics of Astrophysical Discs} (edited by Sellwood,
  J. A.), pp. 199-207. Cambridge Univ. Press, Cambridge.\\
Polyachenko, V. L. and Fridman, A. M. (1972)
  The law of planetary distances.
  {\it Soviet Astr.} {\bf 16}, 123-128.\\
Polyachenko, V. L. and Polyachenko, E. V. (1997) 
  Stability of self-gravitating astrophysical disks.
  {\it J. Exper. Theor. Phys.} {\bf 85}, 417-429.\\
Porco, C. C. (1990) Narrow rings: Observations and theory.
  {\it Adv. Space Res.} {\bf 10}, 221-229.\\ 
Porco, C. C. and Nicholson, P. D. (1987)
  Eccentric features in Saturn's outer C ring.
  {\it Icarus} {\bf 72}, 437-467.\\
Richardson, D. C. (1994) Three code simulations of planetary
  rings. {\it Mon. Not. R. astron. Soc.} {\bf 269}, 493-511.\\
Rosen, P. A. and Lissauer, J. J.
  The Titan-1:0 nodal bending wave in Saturn's Ring C.
  {\it Science} {\bf 241}, 690-694.\\ 
Rosenbluth, M. N., MacDonald, W. M., and Judd, D. L. (1957) 
  Fokker-Planck equation for an inverse-square force.
  {\it Phys. Rev.} {\bf 107}, 1-6.\\
Rukhadze, A. A. and Silin, V. P. (1969) 
  Kinetic theory of drift-dissipative instabilities of a plasma.
  {\it Soviet Phys.-Uspekhi} {\bf 11}, 659-677.\\
Salo, H. (1992) 
  Gravitational wakes in Saturn's rings.
  {\it Nature} {\bf 359}, 619-621.\\
Salo, H. (1995) 
  Simulations of dense planetary rings. III. Self-gravitating
  identical particles.
  {\it Icarus} {\bf 117}, 287-312.\\
Sandel, B. R., Shemansky, D. E., A. L. Broadfoot {\it et al.} (1982) 
  Extreme ultraviolent observations from the Voyager 2 encounter.
  {\it Science} {\bf 215}, 548-553.\\
Schmit, U. and Tscharnuter, W. M. (1995) 
  A fluid dynamical treatment of Saturn's rings.
  {\it Icarus} {\bf 115}, 304-319.\\
Schmit, U. and Tscharnuter, W. M. (1999)
  On the Formation of the Fine-Scale Structure in Saturn's B Ring.
  {\it Icarus} {\bf 138}, 173-187.\\
Sellwood, J. A. and Carlberg, R. G. (1984) 
  Spiral instabilities provoked by accretion and star formation.
  {\it Astrophys. J.} {\bf 282}, 61-74.\\
Shan, Lin-Hua and Goertz, C. K. (1991)
  On the radial structure of Saturn's B ring.
  {\it Astrophys. J.} {\bf 367}, 350-360.\\
Showalter, M. R. (1991) 
  Visual detection of 1981S13, Saturn's eighteenth
  satellite, and its role in the Encke gap.
  {\it Nature} {\bf 351}, 709-713.\\
Showalter, M. R. and Nicholson, P. D. (1990) 
  Saturn's rings through a microscope -- Particle size
  constraints from the Voyager PPS scan.
  {\it Icarus} {\bf 87}, 285-306.\\
Shu, F. H. (1970) 
  On the density wave theory of galactic spirals. II.
  The propagation of the density of wave action.
  {\it Astrophys. J.} {\bf 160}, 99-112.\\
Shu, F. H., Cuzzi, J. N., and Lissauer, J. J. (1983)  
  Bending waves in Saturn's rings.
  {\it Icarus} {\bf 53}, 185-206.\\
Shu, F. H. and Stewart, G. R. (1985) 
  The collisional dynamics of particulate disks.
  {\it Icarus} {\bf 62}, 360-383.\\ 
Shu, F. H., Dones, L., Lissauer, J. J., Yuan, C., and Cuzzi,  
  J. N. (1985) 
  Nonlinear spiral density waves: viscous damping.
  {\it Astrophys. J.} {\bf 299}, 542-573.\\
Sicardy, B. and Brahic, A. (1990)
  The new rings: Contributions of recent ground-based and space
  observations to our knowledge of planetary rings.
  {\it Adv. Space Res.} {\bf 10},  211-219.\\
Smith, B. A., Soderblom, R., R. Batson {\it et al.} (1982) 
  A new look at the Saturn system: the Voyager 2 images.
  {\it Science} {\bf 215}, 504-507.\\
Spahn, F., School, H., and Hertzsch, J.-M. (1994)  
  Structures in planetary rings caused by embedded moonlets.
  {\it Icarus} {\bf 111}, 514-535.\\
Spitzer, L. and Schwarzschild, M. (1953)
  The possible influence of interstellar clouds on stellar 
  velocities. {\it Astrophys. J.} {\bf 118}, 106-112.\\
Sterzik, M. F., Herold, H., Ruder, H., and Willerding, E. (1995)
  Local particle simulations of viscous self-gravitating disks.
  {\it Planet. Space Sci.} {\bf 43}, 259-269.\\
Swanson, D. G. (1989) {\it Plasma Waves}. 
  Academic Press, Boston.\\
Thiessenhusen, Kai-Uwe, Esposito, L. W., Kurths, J., and Shu, F.
  Detection of hidden resonances in Saturn's B ring. (1995)
  {\it Icarus} {\bf 113}, 206-212.\\
Thompson, W. T., Lumme, K., Irvine, W. M., Baum, W. A., and
  Esposito, L. W. (1981) Saturn's rings: Azimuthal variations, 
  phase curves, and radial profiles in four colors.  {\it Icarus.}
  {\bf 46}, 187-200.\\ 
Tremaine, S. (1989) 
  Common processes and problems in disc dynamics.
  In {\it Dynamics of Astrophysical Discs} (edited by Sellwood,
  J. A.), pp. 231-238.  Cambridge Univ. Press, Cambridge.\\
Toomre, A. (1964) 
  On the gravitational stability of a disk of stars.
  {\it Astrophys. J.} {\bf 139}, 1217-1238.\\
Toomre, A. (1977) 
  Theories of spiral structure.
  {\it Annu. Rev. Astr. Astrophys.} {\bf 15}, 437-478.\\
Toomre, A. (1981) 
  What amplifies the spirals?
  In {\it Structure and Evolution of Normal Galaxies}, 
  (edited by Fall, S. M. and Lynden-Bell, D.), pp. 111-136. 
  Cambridge Univ. Press, Cambridge.\\
Tyler, G. L., Marouf, E. A., Simpson, R. A., Zebker, H. A., 
  and Eshleman, R. (1983)
  The microwave opacity of Saturn's rings at wavelengths of 
  3.6 and 13 cm from Voyager 1 radio occultation. 
  {\it Icarus} {\bf 54}, 160-168.\\ 
Ward, W. R. (1981) 
  On the radial structure of Saturn's rings.
  {\it J. Geophys. Res. Lett.} {\bf 8}, 641-643.\\
Willerding, E. (1992) 
  Secular ring instability in the protoplanetary disk.
  {\it Earth, Moon, and Planets} {\bf 56}, 173-192.\\
Wisdom, J. and Tremaine, S. (1988) 
  Local simulations of planetary rings.
  {\it Astron. J.} {\bf 95}, 925-940.\\
Zebker, H. A., Marouf, E. A., and Tyler, G. L. (1985)
  Saturn's rings: Particle size distributions for thin 
  layer model.  {\it Icarus} {\bf 64}, 531-548.\\

\end{document}